%% file: main_NiPS3cluster-Magda.tex
\documentclass[amssymb,twocolumn,floats,amsmath,superscriptaddress]{revtex4-1}
\usepackage{graphicx}
\usepackage{textcomp}
\usepackage{float}
\usepackage{xcolor}
\usepackage[normalem]{ulem}
\usepackage{upgreek}
\usepackage[english]{babel}
\usepackage[utf8]{inputenc}
\usepackage{tikz}
\usepackage{braket}
\usepackage{cleveref}
\usepackage{amssymb}
\usepackage[colorinlistoftodos, color=green!40, prependcaption]{todonotes}
\newcommand\kw{\textcolor{blue}}

\input{preamble}
\def \IFT{Institute of Theoretical Physics, Faculty of Physics, University of Warsaw, Pasteura St. 5, 02-093 Warsaw, Poland}
\def \PW{Department of Semiconductor Materials Engineering Faculty of Fundamental Problems of Technology Wrocław University of Science and Technology Wybrzeże Wyspiańskiego 27, 50-370 Wrocław, Poland}
\def \AC{II. Institute of Physics B and JARA-FIT, RWTH-Aachen University, 52074 Aachen, Germany}
\def \Haifa{Schulich Faculty of Chemistry, Solid State Institute, Russell Berrie Nanotechnology Institute, Technion, Haifa 3200003, Israel}
\def \HaifaMat{Department of Materials Science and Engineering, Technion – Israel Institute of Technology, Haifa 3200003, Israel}
\def \FZJ{Forschungszentrum Jülich, Peter Grünberg Institute (PGI-6),
52425 Jülich, Germany}

\def \FZJ{Forschungszentrum Jülich,
Peter Grünberg Institute (PGI-6),
52425 Jülich, Germany}
\def \Trieste{Physics Department, University of Trieste, Trieste, 34127 Italy}
\def \Elettra{Elettra – Sincrotrone Trieste S.C.p.A., S.S. 14 km 163.5, Trieste, 34149 Italy}
\def \PNNL{Physical and Computational Sciences Directorate and Institute for Integrated Catalysis, Pacific Northwest National Laboratory, Richland, WA 99354, USA}
\begin{document}

\title{Evidence for Many-Body States in NiPS$_3$ Revealed by Angle-Resolved Photoelectron Spectroscopy}
\author{Miłosz Rybak$^*$}\affiliation{\PW}
\author{Benjamin Pestka}\affiliation{\AC}
\author{Biplab Bhattacharyya}\affiliation{\AC}
\author{Jeff Strasdas}\affiliation{\AC}
\author{Adam K. Budniak}\affiliation{\Haifa}
\author{Adi Harchol}\affiliation{\Haifa}
\author{Vitaliy Feyer}\affiliation{\FZJ}
\author{Iulia Cojocariu}\affiliation{\FZJ}\affiliation{\Trieste}\affiliation{\Elettra}
\author{Daniel Baranowski}\affiliation{\FZJ}\affiliation{\PNNL}
\author{Yaron Amouyal}\affiliation{\HaifaMat}
\author{Efrat Lifshitz}\affiliation{\Haifa}
\author{Magdalena Birowska}\affiliation{\IFT}
\author{Markus Morgenstern}\affiliation{\AC}
\author{Krzysztof Wohlfeld}\affiliation{\IFT}
\date{\today}


\def\MM{\textcolor{red}}
\def\kw{\textcolor{blue}}
\def\Jeff{\textcolor{pink}}
\def\Magda{\textcolor{violet}}
\def\Milosz{\textcolor{green}}

\begin{abstract}

We present $\mu$-ARPES spectra of the Mott-insulating van der Waals antiferromagnet NiPS$_3$. Signatures of strong correlations---such as the onset of atomic or atomic–ligand multiplets and spin–orbit–entangled excitons---have been observed in this material by various two-particle spectroscopies, but not previously in photoemission. Our measurements reveal a weakly dispersive feature at the valence-band edge that is absent in DFT+$U$ calculations and remains unchanged across the N\'eel transition. After ruling out multiple alternative interpretations, we show that an exact diagonalization of a NiS$_6$ cluster
yields low-energy final-state configurations of mixed multiplet $d^7$ and $d^8\underline{L}$ character, whose energy differences are consistent with the observed additional feature. This implies that ARPES directly accesses local Ni–S multiplet physics in NiPS$_3$, revealing a many-body structure beyond  mean-field theory. Our results confirm that NiPS$_3$ is an excellent model platform in which strong correlations, reduced dimensionality, and covalent metal–ligand bonding jointly shape both two- and single-particle spectroscopies, underscoring the need for a genuinely quantum many-body description of these types of two-dimensional quantum materials.
\end{abstract}


\maketitle
\noindent {
{$^*$}Corresponding author: M.~Rybak, Email: \href{milosz.rybak@pwr.edu.pl}{milosz.rybak@pwr.edu.pl}} 

\section{Introduction}

\begin{figure*}[!thbp]
\centering
\includegraphics[width=0.85\textwidth]{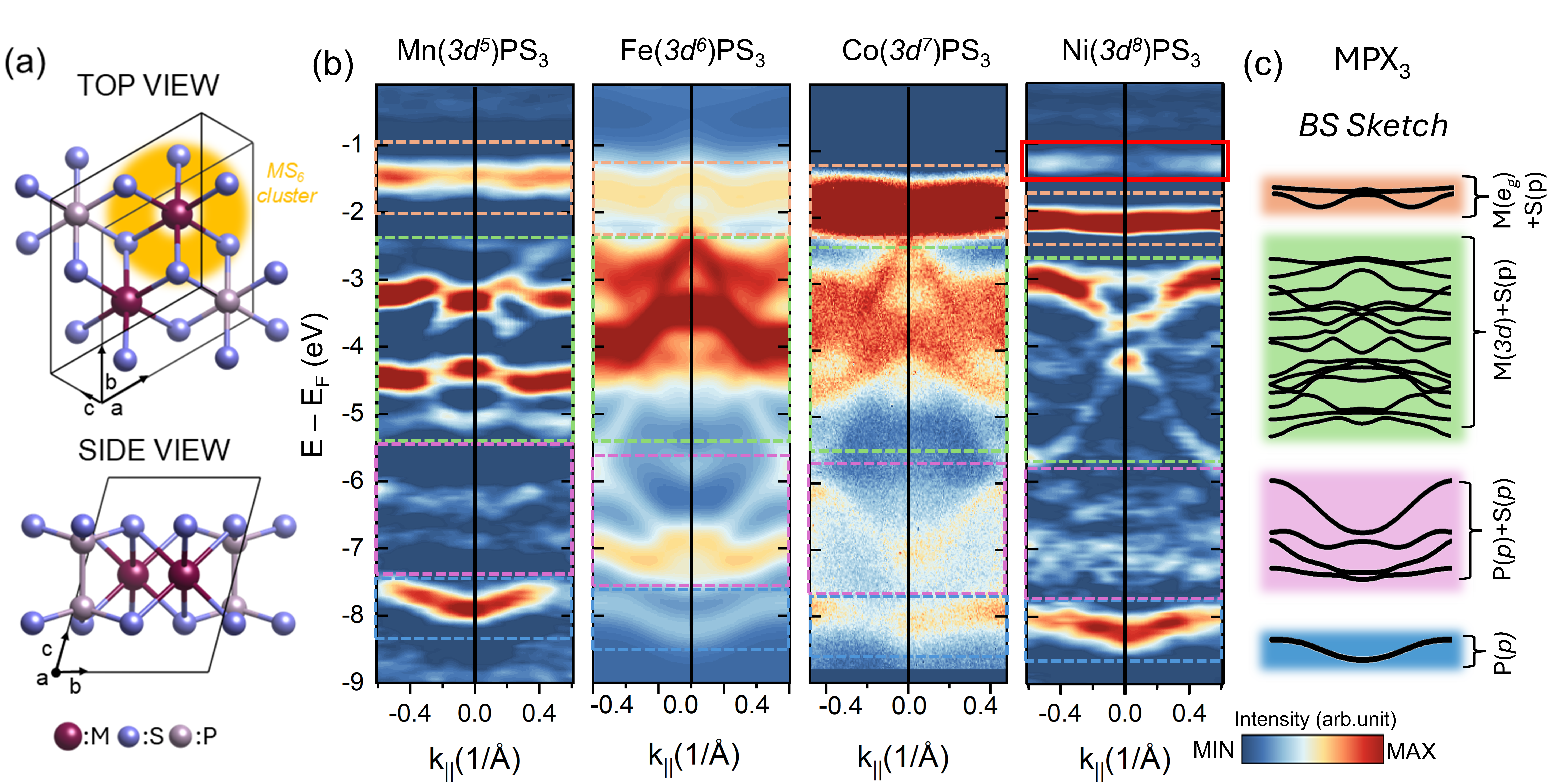}
\vspace{-0.3 cm}
\caption{MPX$_3$ crystal structure and ARPES results:
(a) Atomic structure of MPS$_3$ compounds, illustrating the honeycomb sublattice of MS$_6$ octahedra (top and side views). The yellow-shaded area marks one representative MS$_6$ cluster. (b) Comparison of experimental ARPES curvature plots for MnPS$_3$ \cite{Strasdas2023}, FePS$_3$ \cite{Koitzsch2023}, CoPS$_3$\cite{VOLOSHINA2023140511}\footnote{Figures adapted with permission, respectively, from Refs.~\cite{Strasdas2023,Koitzsch2023,VOLOSHINA2023140511}}, and NiPS$_3$ -- along the M-$\Gamma$-M direction.  The dashed boxes refer to the orbital contribution depicted in (c). 
A characteristic weakly dispersive feature  at low binding energy is highlighted by a red box and is not captured by DFT+$U$ calculations. It emerges only for NiPS$_3$, indicating an additional Ni–S–derived spectroscopic feature.
(c) Schematic summary of the generic DFT+$U$ band structure of all four MPS$_3$ systems. The states at highest binding energy originate predominantly from P $3p$ orbitals  (blue box) followed by a mixture of P and S $3p$ levels (violet box), while the mid-lying manifold arises from hybridized transition-metal $3d$ and chalcogen $3p$ orbitals (green box). The highest-lying states near the Fermi energy correspond to hybridized M ($e_{\rm g}$) levels, whose occupation evolves systematically from Mn ($3d^5$) to Ni ($3d^8$) (orange box).
}
\label{Fig. 1}
\end{figure*}

Layered transition-metal thiophosphates MPS$_3$ (M = Mn, Co, Fe, Ni) have emerged as a versatile platform for exploring correlated magnetism, excitonics, and hybrid metal–ligand physics in van der Waals materials \cite{Wang2018}. Their weak interlayer coupling enables high-quality exfoliation down to the monolayer limit, see Fig. \ref{Fig. 1} (a). At the same time, the transition-metal site furnishes a rich interplay between local Coulomb interactions, ligand-hole configurations, and crystal-field anisotropy \cite{Dedkov_2023}. Within this family, NiPS$_3$ occupies a particularly intriguing position: It lies near the boundary between the charge-transfer and Mott–Hubbard regimes, exhibits substantial Ni–S covalency, and hosts a spin–orbital--entangled ligand-hole exciton \cite{Ho2021, Kang2020, He2024} whose energy, polarization and selection rules are strongly tied to the antiferromagnetic order; this spin–orbital–exciton physics has no direct analogue in MnPS$_3$ \cite{Majchrzak2025, D5TC00322A} or FePS$_3$ \cite{Majchrzak2025,NITSCHKE2025100019}, where optical excitations remain predominantly crystal-field or magnonic in character.

Optical, x-ray absorption spectroscopy (XAS), and resonant inelastic x-ray spectroscopy (RIXS) experiments consistently report sharp excitations attributed to multiplet-split transitions that are predominantly of single-ion Ni$^{2+}$ ($3d^8$) character \cite{Majchrzak2025,Kang2020,He2024}. Interestingly, even
the inherently covalent spin-orbital ligand-hole exciton
shows up as a surprisingly sharp peak in these spectroscopies~\cite{Kang2020, He2024,Song2024, PhysRevLett.131.256504}. This indicates that the strong local correlations found in the Mott-insulating NiPS$_3$ are clearly visible in the two-particle spectroscopic probes.

A central question thus arises: \textit{How does the correlated nature of NiPS$_3$ manifest in a single-particle probe as the angle-resolved photoelectron spectroscopy (ARPES)?} Although ARPES is often interpreted as a direct measurement of the band structure predicted by DFT (including DFT$+U$), in particular, in strongly correlated materials it probes the full one-electron removal spectrum \cite{Hüfner2013}. Such a spectrum has, by definition, a many-body character and can be very different from the itinerant bands obtained from DFT. In fact, one expects that the many-body final states of ARPES on a strongly 
correlated material should show some similarities 
with the multiplet states probed by the aforementioned two-particle probes~\cite{DAMASCELLI2001165}. Consequently, ARPES can reveal spectral features without a direct counterpart in mean-field theory
-- an important point when assessing how reliably ARPES reflects the DFT-derived band structure of correlated materials. A key question for NiPS$_3$ is hence, whether ARPES uncovers additional spectral weight that encodes correlation physics inaccessible to static single-particle approaches, such as DFT.

Previous ARPES studies on MnPS$_3$ \cite{Strasdas2023}, CoPS$_3$ \cite{VOLOSHINA2023140511} and FePS$_3$ \cite{Pestka2024} found good agreement with DFT$+U$ and no anomalous spectral features close to the Fermi level  (Fig.~\ref{Fig. 1}(b)-(c)).
NiPS$_3$ instead revealed a weakly dispersive feature near the valence-band edge that is absent in 
DFT$+U$~\cite{Pestka2025} (red box in Fig.~\ref{Fig. 1}(b)). 
This observation is particularly intriguing in light of optical and XAS studies, which point toward strong metal--ligand hybridization and a multiconfigurational electronic structure in NiPS$_3$, resembling that of other correlated nickel compounds \cite{PhysRevB.30.957}.

In this work, we combine high-resolution $\mu$-ARPES with DFT$+U$ and an exact diagonalization (ED) \cite{PhysRevB.85.165113,WANG2019151}
treatment of the NiS$_6$ octahedron [see Fig.~\ref{Fig. 1}(a)] to clarify this issue. Our ARPES measurements reveal a reproducible, weakly dispersive feature at the top of the valence band that does not appear in mean-field calculations, persists across the antiferromagnetic–paramagnetic transition and shows a nearly identical dispersion as the upper valence band. Considering the full set of experimental conditions together with the theoretical modeling, we find that the behavior of this feature can be explained as a multiplet final-state transition involving hybridized $d^7$ and locally entangled $d^8\underline{L}$~\footnote{i.e. 8 electrons in the nickel $3d$ shell and one ligand hole} configurations. We rule out alternative scenarios such as surface states, defect levels, collective loss features and magnetically dressed quasiparticles leading us to the multiplet-based interpretation. 

Beyond resolving a specific inconsistency between ARPES and DFT$+U$ for NiPS$_3$, our results carry broader implications. They demonstrate that even in structurally simple and nominally insulating systems, ARPES can show features of the many-body structure arising from locally entangled states
that are invisible to conventional band theory, even when the Hubbard $U$ is taken into account on the mean-field level. This underscores the ability of ARPES to act as a sensitive probe for the fully correlated spectral function -- particularly in low-dimensional materials where covalency and reduced screening can stabilise entangled many-body states. This is similar to the well-known case of the cuprates~\cite{Wells1995, Kim1996, Kim2006, Ronning2005, Wang2015, Wrzosek2024, Bacq-Labreuil2025}, 
where one had to go beyond DFT+$U$ to explain ARPES results, but contrasts, e.\,g. to the case of Ca$_2$RuO$_4$ or Sr$_2$IrO$_4$, where despite several theoretical efforts~\cite{Paerschke2017, Klosinski2020, Revenda2025}, ARPES experiments appear to be well explained within the mean-field picture~\cite{Kim2008,  Zhang2013, delaTorre2015, Sutter2017, Louat2019}.

The paper is organized as follows. In Sec. \ref{sec:2} we give an overview of the experimental spectra, beginning with a description of the NiPS$_3$ sample and the ARPES setup in part A, followed by a discussion of the ARPES results concerning the additional feature in part B. Next, in Sec. \ref{sec:3} we present the DFT$+U$ calculations of the NiPS$_3$ band structure. In part \ref{sec:3a}, we discuss the chosen DFT functional and its implementation, in part \ref{sec:3b}, we present the resulting band-structures and its orbital attribution, and in part \ref{sec:3c}, we provide a detailed, intuitive understanding of these results focusing on the spin splitting of the fully occupied Ni t$_{\rm 2g}$ levels and the spin polarization of the ligands.
In Sec. \ref{sec:4} we introduce another computational approach, namely the exact diagonalization of a NiS$_6$ cluster. The method is presented in part \ref{sec:4a}, additional details are given in part \ref{sec:4b}, and the main results are shown in part \ref{sec:4c} comparing the different multiplet final states of the photoemission process and describing its relation to the experimental data. Finally, the paper concludes in Sec. \ref{sec:5} and is supplemented by two appendices.

\section{Experimental results}
\label{sec:2}

\subsection{Material and method}

Details of the experiment are outlined elsewhere \cite{Pestka2025}. In short, NiPS$_3$ crystals were synthesized via the vapor-transport method and subsequently exfoliated onto Si/SiO$_2$ (oxide thickness 90 nm) that is metal-coated with Au (5 nm)/Ti (1 nm). A NiPS$_3$ area showing 15 layers of thickness and a surface roughness of 0.14\,nm was selected by atomic force microscopy for the temperature dependent $\upmu$m-focused angular resolved photoelectron spectroscopy ($\upmu$-ARPES). The magnetic phase transition was confirmed by Raman spectroscopy after the ARPES masurements \cite{Pestka2025,Kuo2016, Muhammad2023, Kim2019b}. ARPES data were recorded at the NanoESCA beamline of the Elettra synchrotron radiation facility in Italy at a background pressure of $5\cdot 10^{-11}$\,mbar~\cite{Schneider2012}. The samples were annealed at 200 $^\circ$C, prior to the $\upmu$-ARPES experiments. The beam spot is $5-10\,\upmu$m and the energy resolution is 50\,meV. The Fermi level is determined on Au. We partially use curvature to display the photoelectron intensity for the sake of better visibility using an  $a_0$ parameter $a_0=0.05$ \cite{Zhang2011}. Related raw data intensity plots are available in \cite{Pestka2025}. Fits of the energy distribution curves $I(E)$ ($I$: Intensity, $E$. energy) as in Fig.~\ref{Fig. 2a}(f)--(j) employ a Shirley background and two Voigt peaks with the energy position, peak intensity and peak widths as fit parameters.

\subsection{ARPES spectra}

\begin{figure}[t!]
\centering
\includegraphics[width=0.45\textwidth]{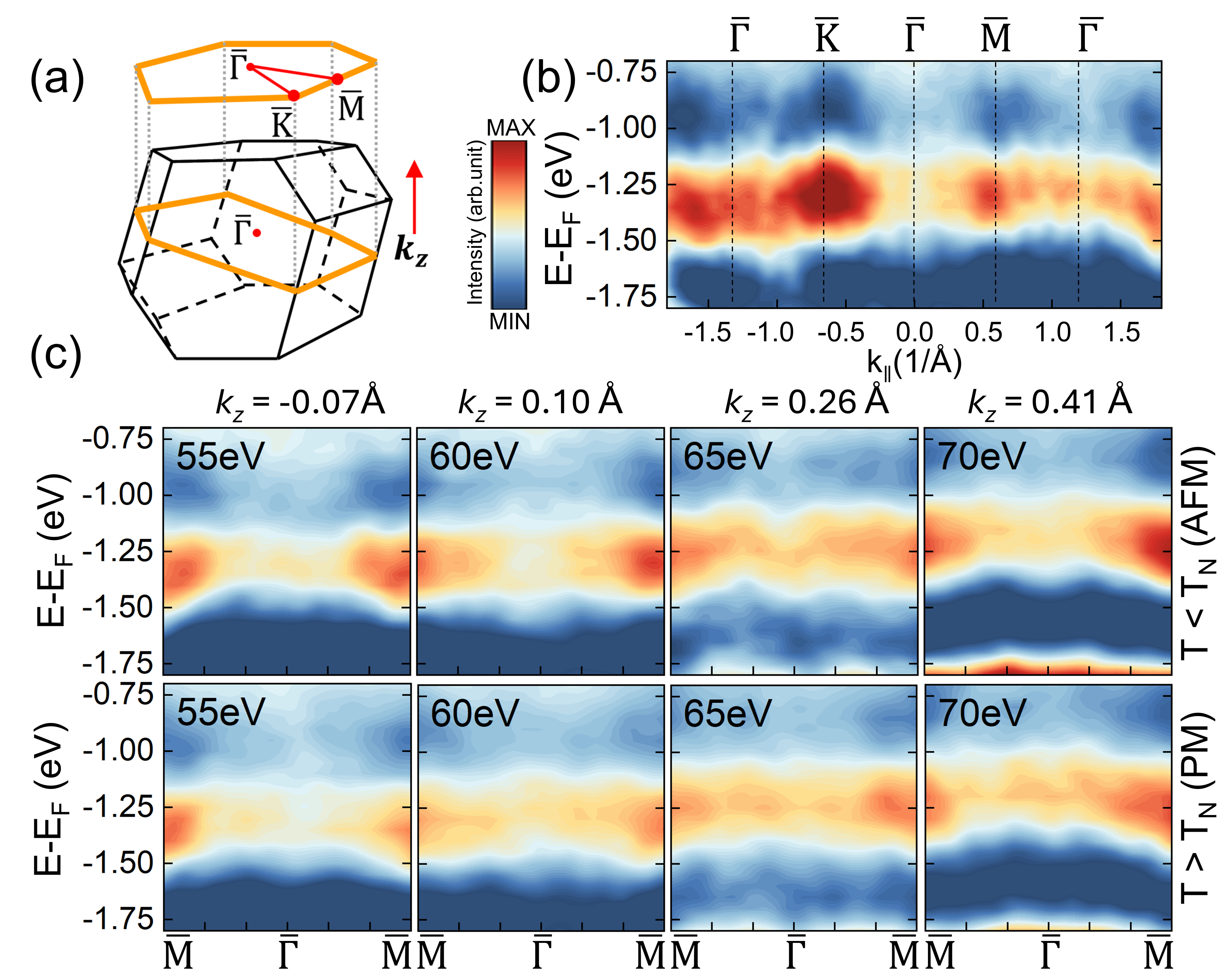}
\vspace{-0.3 cm}
\caption{ (a) Schematic illustration of the Brillouin zone of NiPS$_3$ showing the high-symmetry points and parts of the measured momentum path in (b) within the 2D projection of the Brillouin zone (orange hexagon), in particuar, along $\overline{\Gamma}-\overline{\rm K}$ and $\overline{\Gamma}-\overline{\rm M}$. 
(b) Zoom into the ARPES curvature maps of NiPS$_3$ at T < T$_{\rm N}$ (45\,K) showing the spectral area marked by a red box in Fig.~\ref{Fig. 1}(b). The 
spectral feature centered around –1.3 eV is not captured by DFT+U calculations. It displays a discernable, weak dispersion and a maximum intensity near $\overline{\rm K}$.
(c) Photon-energy-dependent ARPES curvature plots collected for h$\nu$ = 55–70\,eV roughly corresponding to the different k$_z$ values as displayed on top \cite{Pestka2025}.  A distinct variation of the band position and curvature appears indicating 
a dispersion along all three momentum directions. Below and above the Néel temperature (upper and lower row), the dispersion and strength of the spectral feature remains unchanged within experimental accuracy. This indicates that the electronic structure associated with this spectral feature is largely insensitive to the magnetic ordering.
}
\label{Fig. 2}
\end{figure}

\begin{figure*}[t!]
\centering
\includegraphics[width=0.9\textwidth]{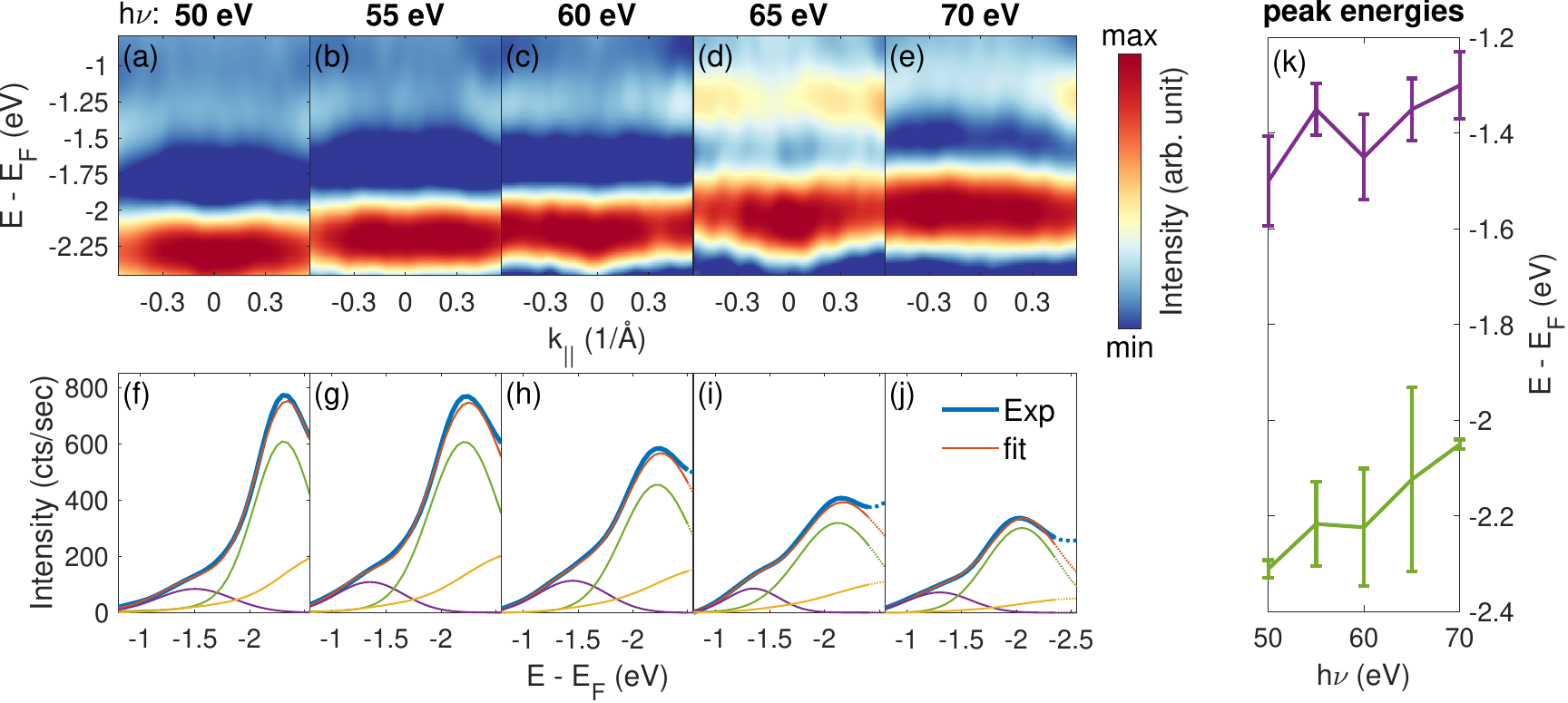}
\vspace{-0.3 cm}
\caption{(a)--(e) ARPES curvature plots along $\overline{\rm M}-\overline{\Gamma}-\overline{\rm M}$ for several photon energies $h\nu$ as marked on top, $T=40$\,K. (f)--(j) ARPES intensity as function of energy for $k_\parallel=0$ and the  $h\nu$ marked at the curvature plots above. The experimental data  (blue lines) are fitted by two Voigt peaks (green and violet lines) and a Shirley background (orange line) leading to the red line (see text). (k) Peak positions of the two Voigt peaks of (f)--(j) with error bars.}
\label{Fig. 2a}
\end{figure*}

Most importantly, the $\mu$-ARPES measurements reveal an additional, weakly dispersive feature at the top of the valence band of NiPS$_3$ that is absent for other MPS$_3$ compounds (Fig.~\ref{Fig. 1}(d), Fig.~\ref{Fig. 2}(b)). 
Revisiting the ARPES datasets from earlier studies \cite{Pestka2025, Pestka2024, Strasdas2023} led us to the following four essential results relevant for the current study: 
\begin{enumerate}
    \item Generally, there is a very good agreement between DFT$+U$ band structure and the valence band dispersion measured by $\upmu$-ARPES for MnPS$_3$, FePS$_3$ and NiPS$_3$ (Fig.~\ref{Fig. 1})(b)-(c)), if one adequately  takes  the (simplified) selection rules of the photoemission process  into account \cite{Moser2017}.
    \item  Only for NiPS$_3$, we persistently  find an additional shoulder-like, weakly dispersive feature above the valence-band maximum (red box in Fig.~\ref{Fig. 1}(b)) that is absent in all DFT$+U$ calculations. 
    \item The 3D $\boldsymbol{k}$-space dispersion of this  additional shoulder largely replicates the dispersion of the upper valence band that exhibits a spin-polarized antibonding Ni 3d$_{\rm e_{\rm g}}$/S 3p character (Fig.~\ref{Fig. 1}(c), Fig.~\ref{fig:covalency}, section~\ref{sec:3ccc}). We find that the shoulder is consistently roughly 700\,meV above the upper valence band (Fig.~\ref{Fig. 2a}).
    \item For the shoulder, neither the dispersion  nor its distance to the upper valence band is influenced by the magnetic order. i.\,e. it is identical above and below $T_{\rm N}$ (Fig.~\ref{Fig. 2}(c)).  
\end{enumerate}
 
 To explain the additional shoulder, we start by discussing possible extrinsic or artefactual origins.

Firstly, surface-related states can be ruled out, since the feature shows $k_z$ dispersion (Fig.~\ref{Fig. 2}(c), \ref{Fig. 2a}(k)). 
Its behavior therefore corresponds to a bulk-derived state, not to a surface state or a surface resonance \cite{Hüfner2013, Damascelli2003}.

Secondly, substrate-induced states from the Au/Ti covered substrate are unlikely. Bands from Au with little dispersion are d-levels at energies below $E-E_{\rm F}=-2$\,eV in contrast to the band-like feature at $E-E_{\rm F}=-1.3$\,eV observed here. Titanium is well buried below the Au as confirmed by x-ray photoelectron spectroscopy \cite{Pestka2025} such that it cannot contribute to the ARPES data.  


Thirdly, defect-induced states are improbable. The XPS data did not show any foreign components \cite{Pestka2025}. 
Moreover, it is unlikely that a defect state shows dispersion and even follows the dispersion of an identified valence band of intact NiPS$_3$.

Fourthly, collective-loss features of the photolectrons such as by plasmons typically produce broad, photon-energy–tracking replicas rather than a narrow, dispersive band. Moreover, as loss features they must appear at the low energy side of the electronic single-particle bands observed by ARPES in contrast to the experiment. 

Finally, magnetically dressed quasiparticles, i.\,e. spin polarons~\cite{Martinez1991}, can also be ruled out, since the feature shows no measurable change across $T_{\rm N}$ (Fig.~\ref{Fig. 2}(c)). The absence of spin polarons can be understood by noting that the ratio of magnetic and kinetic energy in NiPS$_3$ differs substantially from those in the cuprates, which exhibit clear and well-established signatures of spin polarons~\cite{Wells1995, Ronning2005, Wang2015, Wrzosek2024, Bacq-Labreuil2025}. In the cuprates, the spin exchange $J$ is nearly half of the effective copper--copper hopping $t$. This leads to a non-negligible spin-polaron spectral weight which scales as $\propto J/$ t \cite{Martinez1991}.
By contrast, in NiPS$_3$ the ratio of the largest (in fact, third-neighbor) exchange interaction $J$ to the nickel--nickel hopping $t$ is of order $1/10$~\cite{Autieri2022}. As a result, the spin-polaron spectral weight is strongly suppressed and merges into the incoherent background.

Having systematically excluded surface/interface bands, collective losses, magnetic polarons and largely also defect levels, the remaining interpretation is that the additional spectral weight derives from a localized, multiplet-split final-state transition of the NiS$_6$ cluster. Its very similar dispersion to the top valence band additionally indicates a common initial state. Moreover, its insensitivity to the AFM--PM transition indicates predominantly local physics independent of the long-range magnetic order. The fact that DFT$+U$ reproduces all other bands but not this one reinforces the picture: mean-field band theory cannot encode multiplet splitting or ligand-hole mixtures, e.\,g. in the final state---precisely the ingredients which might generate such an additional spectral feature.

Hence, we complement the experimental analysis with DFT$+U$, establishing the single-particle baseline, by using a multiplet-resolved ED for the NiS$_6$ cluster. This naturally produces a low-energy final-state multiplet that matches the measured energy separation between the shoulder and the upper valence band. Together, experiment and theory consistently support the conjecture that the additional spectral shoulder in NiPS$_3$ originates from a multiplet-split final-state caused by its correlated electronic structure.

\begin{figure}[t!]
\centering
\includegraphics[width=0.5\textwidth]{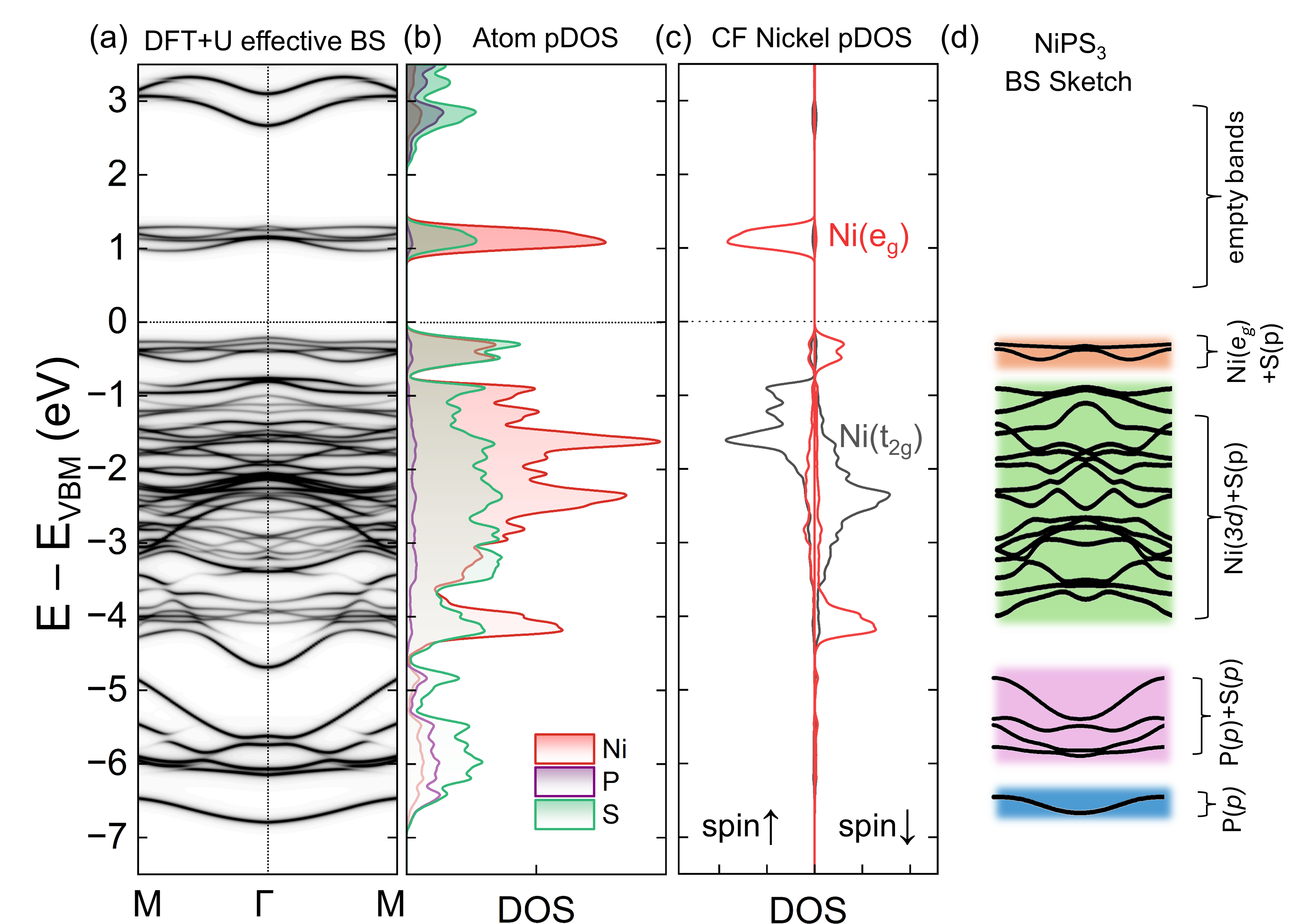}
\vspace{-0.3 cm}
\caption{(a) Effective band structure of NiPS$_3$ obtained from DFT+$U$ calculations and unfolded onto the primitive (non-magnetic) Brillouin zone to enable direct comparison with the experimental spectra. The details of the unfolding procedure and its justification are provided in \cite{Pestka2025}. $E_{VBM}$ denotes the valence band maximum.
(b) Atom-projected density of states (pDOS) for Ni, P, and S, showing the dominant Ni–S hybridization within the valence manifold.
(c) Crystal-field-resolved Ni 3$d$ pDOS highlighting the separation between $t{_{2g}}$ and $e_{\rm g}$ states, with the latter forming the upper valence band.
(d) Simplified schematic of the NiPS$_3$ band structure, illustrating the correspondence between the calculated DFT+$U$ results and the characteristic arrangement of metal $e_{\rm g}$/$t_{\rm 2 g}$ levels and ligand-derived bands.
}
\label{Fig. 3}
\end{figure}

\section{DFT$+U$ approach}
\label{sec:3}
\subsection{Computational details}
\label{sec:3a}
The DFT calculations were performed using the generalized gradient approximation within the PBE flavor \cite{PhysRevLett.77.3865}, as implemented in the VASP software \cite{Kresse1996}. The ion–electron interactions were described by the projector augmented wave (PAW) method  \cite{Holzwarth2001}. Plane-wave basis cutoff and $\Gamma$ centered Monkhorst-Pack \cite{Monkhorst1976} k-point grid were set to 550 eV and $8\times14\times8$, respectively. A Gaussian smearing of 0.05 eV was employed for the Brillouin zone (BZ) integration. The interlayer vdW forces were treated within Grimme scheme using D3 correction \cite{doi:10.1063/1.3382344}.

All calculations were performed within the rotationally invariant DFT$+U$ formalism of Liechtenstein {\it et al.}~\cite{PhysRevB.52.R5467}. Whenever no explicit value of the Hund's exchange $J_H$ is stated, the calculations were carried out with $J_H= 0.0$ eV, in which case the Liechtenstein scheme becomes equivalent to the Dudarev formulation~\cite{PhysRevB.57.1505}.
The position of the atoms and unit cell were fully optimized within the PBE+$U$ approach. Moreover, we used the widely employed band unfolding method to obtain an effective electronic structure~\cite{Popescu2012}.

\subsection{Results}
\label{sec:3b}

\begin{figure*}[!thbp]
\centering
\includegraphics[width=1\textwidth]{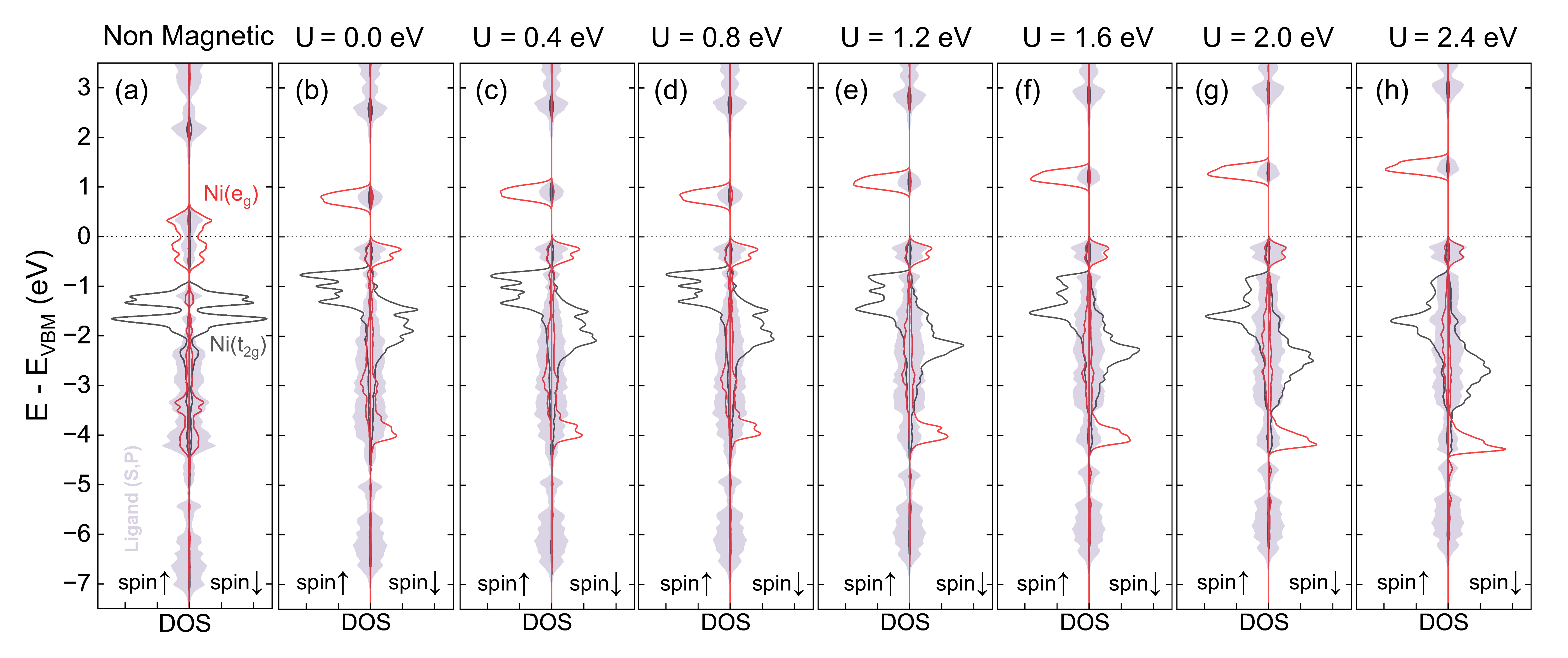}
\vspace{-0.3 cm}
\caption{Evolution of the spin-dependent DOS with an increasing value of Hubbard $U$ in the DFT$+U$ approach. Calculations performed for NiPS$_3$ at $J_H=0.0$ eV, see text for further details. The calculated spin-dependent DOS is additionally projected to the Ni e$_{\rm g}$ orbitals (red), the Ni t$_{\rm 2g}$ orbitals (black) and the s and p orbitals of the ligand atoms S and P (lilac grey). Note that the left-most panel is obtained for $U=0.0$ eV and a non-magnetic ground state.
}
\label{fig:evolutionU}
\end{figure*}

To understand the electronic properties and correlation effects in these materials, we compared our measured ARPES spectrum of NiPS$_3$ with both theoretically predicted spectra and previously reported experimental spectra of related compounds. Formally, in the most simple approximation, modeling the photoemission spectrum requires computing the single-particle spectral function of the system, which is directly related to the imaginary part of the retarded single-particle Green’s function $A(\mathbf{k},\omega) = -\frac{1}{\pi} \operatorname{Im} G(\mathbf{k},\omega).$ Moreover, in the non-interacting limit, the imaginary part of the Green's function is merely a set of
delta functions with peaks positioned at 
$\omega = \varepsilon_n(\mathbf{k}) $ with $\varepsilon_n(\mathbf{k})$
being the energy of the $n$-th band obtained e.g. from DFT (see Sec. IVA for more details). For materials where electron-electron interactions are present but do not entirely dominate their behavior, methods such as the Hubbard correction (DFT$+U$) or the inclusion of a fraction of Hartree-Fock exchange within hybrid functionals (e.g., HSE06) are often sufficient to capture most of the ARPES spectrum -- that mostly contains the interaction-renormalised bands. The literature confirms the applicability of these approaches for MnPS$_3$ \cite{Strasdas2023}, FePS$_3$ \cite{Pestka2024}, and CoPS$_3$ \cite{VOLOSHINA2023140511}. However, in the case of 
NiPS$_3$, neither DFT$+U$ nor hybrid functionals were able to fully reproduce the band structure~\cite{Pestka2025}.

By comparing our theoretical results across different members of the MPS$_3$ family (M: transition metal), we observe that the overall band sequence remains rather similar (Fig.~\ref{Fig. 1}b--c). 
At large binding energies ($\sim$7 eV below the valence band maximum), a well-separated phosphorus-derived band appears (blue area in Fig.~\ref{Fig. 1}c). Above it, four hybridized phosphorus and sulfur bands are found (violet area in Fig.~\ref{Fig. 1}c), followed by a manifold of sulfur-derived states that strongly hybridize with the transition metal \textit{d}-states (green area in Fig.~\ref{Fig. 1}c). This region exhibits the most significant variations between different materials, as it hosts the $3d$ states of the transition metal ions, whose filling varies across the series.

Slightly above this manifold, we identify a set of relatively weakly dispersive $e_{\rm g}$ states that are hybridized with S 3p states. Their number depends on the long-range magnetic order, i.\,e. the size of the magnetic unit cell (orange area in Fig.~\ref{Fig. 1}c). In the Néel-ordered phase, two such states appear, while in the zigzag phase, four states are observed, due to the doubling of the unit cell imposed by this magnetic symmetry. However, in the case of NiPS$_3$, we detect an additional, weakly dispersive band  (red box in Fig.~\ref{Fig. 1}b, NiPS$_3$,  that cannot be reproduced using either DFT$+U$ or hybrid functionals. This suggests that the mean-field approximation is inadequate and that electronic correlations in NiPS$_3$ are significantly stronger than in the other members of this family.

\begin{figure*}[!thbp]
\centering
\includegraphics[width=1\textwidth]{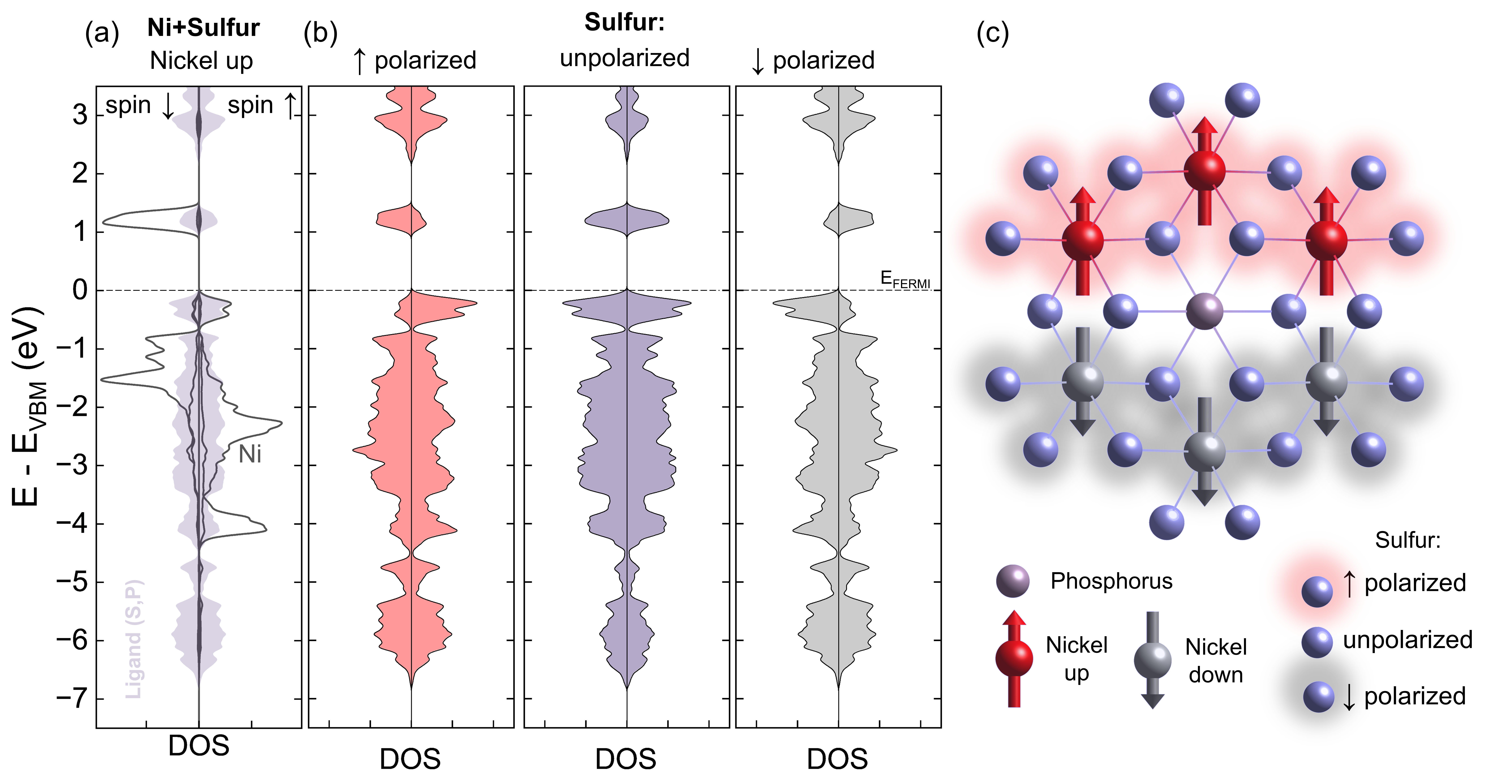}
\vspace{-0.3 cm}
\caption{Spin-dependent DOS in the DFT$+U$ approach: (a)  nickel (black) and sulfur (lilac grey) orbitals  [equivalent to the panel at $U_{\rm H}=1.6$\,eV in  Fig.~\ref{fig:evolutionU}];
(b) only sulfur orbitals, showing DOS for spin-up
polarized (left panel), unpolarized (middle panel),
and spin-down polarized (right panel) sites as depicted in (c);
(c) Cartoon showing polarization of sulfur atoms depending on the location with respect to the 
spin-polarized nickel atoms in the zigzag AF-ordered state.
Calculations performed for NiPS$_3$ at $U=1.6$ eV
and $J_H=0.0$ eV, see text for further details. 
}
\label{fig:covalency}
\end{figure*}

\subsection{Discussion:\\
Dependence on Hubbard $U$ and Hund's exchange $J_H$}
\label{sec:3c}
In what follows we derive a qualitative understanding of the central structures of the NiPS$_3$ spin-dependent sublattice-projected DOS obtained within DFT$+U$.
In particular, there exist two main features of the obtained DOS that need to be understood: (i) the splitting of the $e_{\rm g}$ bands into three spin-projected energy regions of DOS (one unoccupied for one type of spin and two occupied for the other spin (Fig.~\ref{Fig. 3}c)); (ii) the splitting of the occupied $t_{\rm 2 g}$ band into two spin-projected DOS (both occupied (Fig.~\ref{Fig. 3}c)).

\subsubsection{Splitting of the $e_{\rm g}$ bands into three spin-projected DOS.}
\label{sec:3ccc}
We begin the discussion by recalling the textbook results concerning the canonical single-band Hubbard model at half-filling on the bipartite lattice. In this case a mean-field treatment of the Hubbard interaction $U$ splits the half-occupied, doubly spin-degenerate, band crossing the Fermi level into two Hubbard subbands. This phenomenon is directly visible in the sublattice-projected DOS. For example, for the DOS projected onto the so-called spin-up sublattice, we obtain 
a spin-up occupied DOS below the Fermi level and a spin-down unoccupied DOS
above the Fermi level. This phenomenon is also well-visible in the DFT$+U$ calculations on materials which essentially have one half-occupied band crossing the Fermi level. 

The above description already partially explains the observed splitting of the $e_{\rm g}$ bands. Here, as a result of the Hubbard $U$ on Ni the half-filled quadruple-degenerate $e_{\rm g}$ band splits into an unoccupied spin-polarized~\footnote{Here, and in what follows, by spin-polarized bands we mean the onset of spin polarization in the sublattice-projected DOS.} double-degenerate $e_{\rm g}$ subband and an occupied spin-polarized double-degenerate $e_{\rm g}$ subband at low energies (around -4 eV), see Fig.~\ref{fig:evolutionU}.

Next, one needs to explain the additional $e_{\rm g}$ subband that is always occupied, shows spin polarization, and is located close to the Fermi level. 
As Fig.~\ref{fig:evolutionU} shows this feature rather weakly depends on the Hubbard $U$. Its spectral weight only weakly goes down with increasing $U$. This suggests that it cannot originate  {\it directly} from the Hubbard $U$ on the nickel site. In fact, we can track the splitting of the occupied states as stemming from a nickel-sulfur covalency.

A more microscopic understanding of the low-energy Ni-derived states within the DFT+$U$ description can be obtained by analyzing the bonding–antibonding nature of the $e_{\rm g}$ manifold, as discussed in detail in the Appendix~\ref{antibonding}. Here we summarize the essential physical picture: 
The occupied Ni $e_{\rm g}$-projected density of states exhibits two distinct features, which we refer to as the \emph{upper $e_{\rm g}$} and \emph{lower $e_{\rm g}$} states. The lower $e_{\rm g}$ feature appears at higher binding energy and is predominantly Ni-centered, whereas the upper $e_{\rm g}$ feature lies closer to the valence-band maximum and carries a more substantial sulfur contribution (Fig.~\ref{fig:evolutionU}). This two-peak structure reflects strong $\sigma$-type hybridization between Ni $e_{\rm g}$ orbitals, which point directly toward the surrounding sulfur ligands, and ligand $p_\sigma$ states.
Within this picture, the lower $e_{\rm g}$ feature can be identified as a bonding combination of Ni $e_{\rm g}$ and S $p$ orbitals, while the upper $e_{\rm g}$ feature corresponds to the associated antibonding state. The strong directional overlap responsible for this splitting is specific to the $e_{\rm g}$ symmetry; the Ni $t_{\rm 2 g}$ orbitals, which point between the ligands, hybridize much more weakly and therefore do not form an equally well-resolved bonding–antibonding pair.
The remaining minor evolution of these two $e_{\rm g}$ features with increasing Hubbard interaction $U$ further supports this interpretation. As discussed in the Appendix~\ref{antibonding}, increasing $U$ systematically reduces the spectral weight of the upper $e_{\rm g}$ state near the valence-band edge, while enhancing the weight of the lower $e_{\rm g}$ feature at higher binding energies. This redistribution reflects a gradual reduction of Ni–S covalency as electronic correlations push the Ni $3d$ levels away from the ligand manifold and favor a more localized Ni-centered electronic structure.

\begin{figure*}[!thbp]
\centering
\includegraphics[width=1\textwidth]{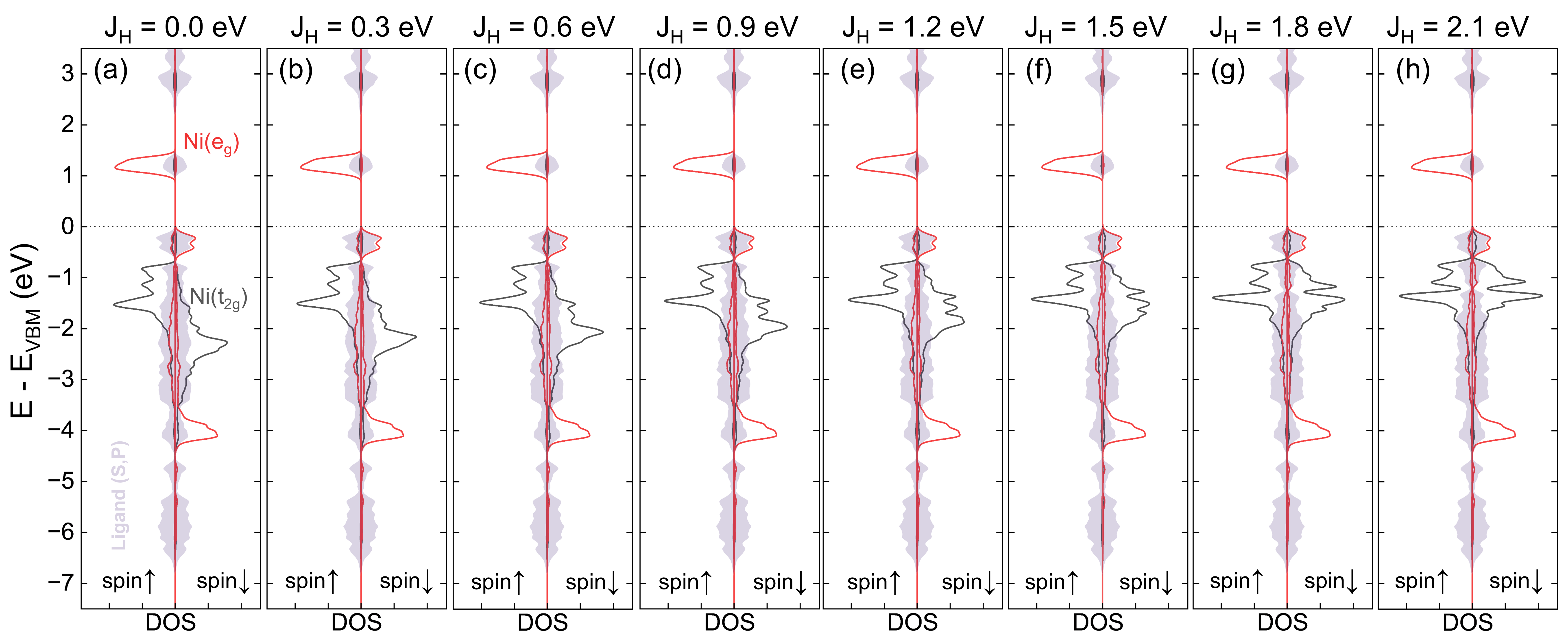}
\vspace{-0.3 cm}
\caption{Evolution of the spin-dependent DOS with an increasing value of Hund's exchange $J_H$ in the DFT$+U$ approach. Calculations performed for NiPS$_3$ at $U=1.6$ eV, see text for further details.}
\label{fig:evolutionJH}
\end{figure*}

Additional insight is provided by the behavior of the local magnetic moments of the ligands. 
Figure~\ref{fig:covalency}(a) shows that the spin-polarized sulfur orbitals just below the Fermi level exhibit the nearly identical DOS as the Ni orbitals of the same spin $\uparrow$ direction. Moreover, the ligands show a substantial spin polarization, if they have two Ni neighbors with the identical spin direction (Fig.~\ref{fig:covalency}b--c). The spins of these Ni atoms and the S atom in between are aligned.
In contrast, if the two neighboring Ni atoms have opposite spin, the ligand does not exhibit a spin polarization as usual, e. g., if the ligand just mediates an antiferromagenetic superexchange coupling, see \mbox{Fig.~\ref{fig:covalency}(b)--(c)}.

We now show how the above-revealed magnetisation supports our bonding–antibonding interpretation of the $e_{\rm g}$ manifold. For that, we note that it is only the upper $e_{\rm g}$ state, owing to its significant ligand character, that induces a finite spin polarization on the sulfur atoms neighboring the magnetic Ni ions (Fig.~\ref{fig:covalency}(b)). As the upper $e_{\rm g}$ weight is suppressed with increasing Hubbard $U$, this ligand polarization is correspondingly reduced, while the magnetic moment on the Ni site increases. These opposite trends provide a clear fingerprint of diminishing metal–ligand hybridization and are fully consistent with the bonding–antibonding interpretation of the occupied $e_{\rm g}$ manifold.
Finally, the robustness of this picture is corroborated by comparing different members of the MPX$_3$ family. In compounds with a large crystal-field splitting, such as MnPS$_3$ and NiPS$_3$, the antibonding upper $e_{\rm g}$ state remains well separated and predominantly of $e_{\rm g}$ symmetry \cite{Pestka2025}. In contrast, in FePS$_3$ and CoPS$_3$, where the crystal-field splitting is reduced, both $e_{\rm g}$ and $t_{\rm 2 g}$ states participate in the antibonding feature, leading to mixed orbital character (See detailed analysis in Appendix~\ref{antibonding}). This systematic trend further supports the interpretation of the upper and lower $e_{\rm g}$ features as bonding and antibonding states arising from metal–ligand hybridization as captured within a mean-field description.

Altogether, this brings us to the conclusion that the spin polarized $e_{\rm g}$ band just below the Fermi level originates in the following manner: first, the spin-polarized $e_{\rm g}$ orbital at low energy of around $-4$ eV (that stems from the standard Hubbard-$U$ mean-field polarization) hybridise with neighboring ligands, polarises them and forms a bonding state strongly localized at the Ni site; second, the spin-polarized ligands also form an antibonding state with the neighboring $e_{\rm g}$ orbitals and this state is at much higher energy of about $-0.5$ eV and localized equally at the S and the Ni site. Note that the same type of splitting into bonding and antibonding states occurs for the other spin level in the unoccupied conduction band. At $+1$\,eV, the S states are strongly spin polarized, in particular, if they bridge two Ni atoms with the same spin direction [Fig.~\ref{fig:covalency}(a)--(b)]. The bonding state at higher energy is not displayed.

\subsubsection{Splitting of the $t_{\rm 2 g}$ band into two spin-projected DOS.}

Next, we move to explaining the spin-polarization splitting observed within the fully occupied \textit{t}$_{2g}$ bands. Such a splitting cannot be understood within a simple picture of occupied versus unoccupied states and therefore requires going beyond a single-orbital description. Our results show, however, that this splitting is again a covalency-driven effect, closely related to the mechanism discussed above for the \textit{e}$_g$ states.

We first note that the magnitude of the \textit{t}$_{2g}$ spin splitting increases on average with increasing Hubbard $U$, in close analogy to the behavior of the \textit{e}$_g$ bands, see Fig.~\ref{fig:evolutionU}. In addition, the correspondingly projected density of states broadens for both spin channels. The former already suggests that the \textit{t}$_{2g}$ splitting is not intrinsic to the \textit{t}$_{2g}$ shell itself, but is instead induced by the strong spin polarization of the half-filled \textit{e}$_g$ manifold.

Having established that the spin polarization of the \textit{e}$_g$ states acts as a driver for the \textit{t}$_{2g}$ splitting, it is instructive to analyze its dependence on Hund’s exchange $J_H$, see Fig.~\ref{fig:evolutionJH}. The parameter $J_H$ quantifies the intra-atomic exchange interaction favoring parallel spin alignment across different orbitals of the same atomic shell and thus controls the strength of the effective Hund-driven exchange field acting between the \textit{e}$_g$ and \textit{t}$_{2g}$ manifolds (see Appendix~\ref{Mean-field}). 

While the positions of the \textit{e}$_g$ features are found to be almost insensitive to variations of $J_H$, the spin splitting of the fully occupied \textit{t}$_{2g}$ bands is strongly reduced upon increasing $J_H$. Importantly, this reduction does \emph{not} originate from a suppression of the \textit{e}$_g$ or ligand spin polarization, which remains robust over the entire range of $J_H$ considered. Instead it basically follows from the fact that with increasing $J_H$ it becomes more and more energetically favorable for the $t_{\rm 2g}$ electrons to have their spins aligned with the majority spin of the $e_{\rm g}$ electrons. This is demonstrated explicitly by deriving Eq.~\eqref{eq:t2g_split} in Appendix~\ref{Mean-field} -- which basically follows from the Hubbard-Kanomori interaction (with the so-called double counting correction also playing an important role, see Appendix~\ref{Mean-field} for more details). Here it is important to note that according to Eq.~\eqref{eq:t2g_split} 
the spin up (minority in the presented results of Fig~\ref{fig:evolutionJH}) $t_{\rm 2g}$
electrons acquire a more positive potential $\propto -J_H M_{e_{\rm g}}$
(since $M_{e_{\rm g}} <0$ in our convention of of Fig.~\ref{fig:evolutionJH}).
Thus their energy shifts downwards with increasing $J_H$, exactly as observed in  Fig.~\ref{fig:evolutionJH}.
On the other hand, the spin down (i.e. majority) $t_{\rm 2g}$ electrons acquire a negative potential 
$\propto J_H M_{e_{\rm g}}$
(remember that $M_{e_{\rm g}} <0$
in our convention) and shift upwards with increasing $J_H$, 
in agreement with Fig.~\ref{fig:evolutionJH}.

Last but not least, it is also important to note that a reversal phenomenon does not happen for the $e_{\rm g}$ electrons -- for the $t_{\rm 2g}$ electrons are fully filled, see Appendix~\ref{Mean-field}
for details.



\subsubsection{Summary}

To sum up, the DFT+$U$ analysis provides a fully consistent mean-field understanding of the electronic structure of NiPS$_3$. The overall band structure is governed by the crystal-field splitting between the $e_{\rm g}$ and $t_{\rm 2 g}$ manifolds, strong Ni–S covalency, and the resulting bonding–antibonding splitting of the $e_{\rm g}$ states. The evolution of the spin-dependent density of states with both $U$ and $J_H$ is well understood: Hubbard interactions polarize the $e_{\rm g}$ shell, covalent hybridization transfers this polarization to the ligands, and Hund’s exchange controls the induced spin splitting of the occupied $t_{\rm 2 g}$ bands by lowering the binding energy of its majority-spin states. This picture is internally consistent. It is supported by systematic trends of the hybridization across the MPX$_3$ family, as demonstrated in Fig.~\ref{fig:s3} of the Appendix~\ref{antibonding}.
Within this mean-field framework, all prominent features of the calculated band structure are naturally explained, and no additional single-particle band is expected near the top of the valence band. While this description successfully captures the overall dispersion, orbital character, and magnetic trends observed in ARPES, it also makes clear that the experimentally observed weakly dispersive feature cannot originate from mean-field band physics. Its absence in DFT+$U$ therefore points 
to physics beyond the single-particle picture, motivating the use of a complementary many-body cluster approach in the following.

\begin{figure*}[!thbp]
\centering
\includegraphics[width=1\textwidth]{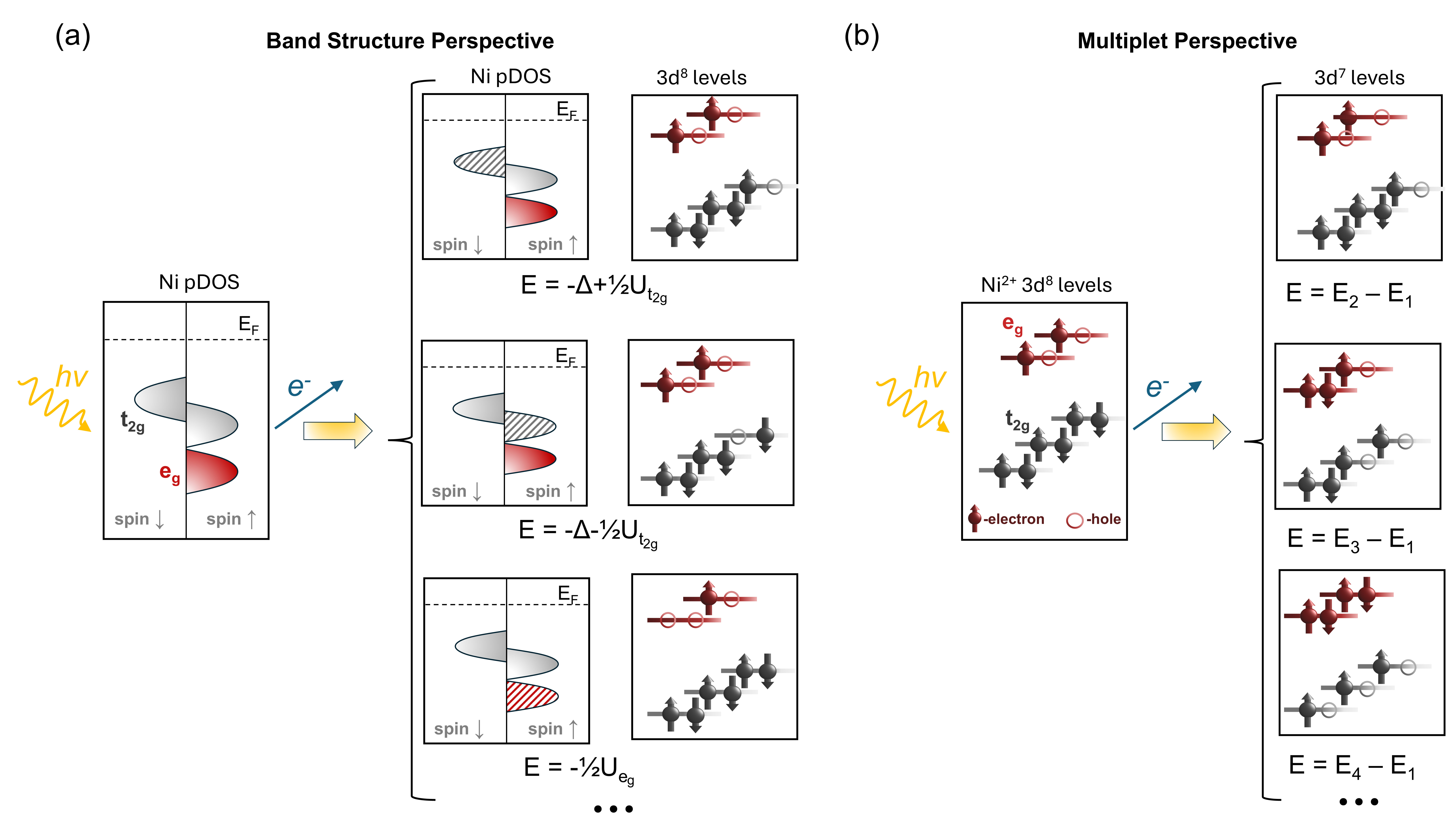}
\vspace{-0.3 cm}
\caption{
Cartoon explaining two approaches to the calculation of the theoretical ARPES spectrum:
(a) Using the DFT$+U$ approach to model ARPES. Left: the three dominant features of the Ni$_\uparrow$-sublattice-projected DOS. Right: a hole occupying one of these three dominant DOS features leads to three ARPES features located at energies that correspond (in the linear response approximation) to their location in the DFT$+U$ calculations of the undoped limit. (b) In the cluster approach it is assumed that ARPES probes energy levels of a NiS$_6$ cluster. Left: the two energy levels of the Ni$^{2+}$ ion in a dominant cubic environment (hybridising ligand $S$-states skipped for clarity). Right: removing one electron by ARPES from the Ni$^{2+}$ ion leads to three dominant `multiplet' eigenstates of the Ni$^{3+}$ ion  with three eigenergies $E_2$, $E_3$ and $E_4$ that correspond to the energy of ARPES features in this approach. 
}
\label{fig:cartoon}
\end{figure*}

\section{Cluster approach}
\label{sec:4}

\subsection{General remarks}
\label{sec:4a}

The discussion in the previous section has shown that the ARPES spectrum calculated using the DFT$+U$ approach cannot explain the spectral feature closest to $E_{\rm F}$ that exhibits the same dispersion as the assigned single-particle band below and is observed across the antiferromagentic-paramagentic transition. 
Therefore, in what follows, we employ another method -- the so-called cluster approach 
which is complementary to DFT$+U$. The fundamnetal difference between the two approaches is explained in detail below and illustrated in Fig.~\ref{fig:cartoon}.

Generally, the DFT$+U$ approach is often adequate to describe ARPES data in weakly correlated band insulators. In contrast, a cluster approach can be necessary in strongly correlated, charge--transfer systems such as NiPS$_3$. In weakly interacting materials, electrons are well described as delocalized Bloch waves and the electronic structure is naturally formulated in momentum space (Fig.~\ref{fig:cartoon}(a)). In this limit, the electronic Hamiltonian is effectively diagonal in a single-particle basis and ARPES directly maps the dispersion of these eigenstates. A cluster-based description, see Fig.~\ref{fig:cartoon}(b), is neither required nor meaningful as the relevant physics is governed by extended bands rather than local degrees of freedom.
NiPS$_3$ lies in the opposite limit. Strong local Coulomb interactions combined with substantial Ni--S covalency lead to the localization of electrons on NiS$_6$ units. As a consequence, the appropriate low-energy degrees of freedom are not independent single-particle Bloch states, but many-body eigenstates of coupled transition-metal–ligand clusters. ARPES can then be interpreted as a probe of local electron-removal excitations, only weakly dressed by intersite propagation.

In the conventional, linear-response description of ARPES the photocurrent measured in ARPES is proportional to the one-particle electron-removal spectral function
$A(\mathbf{k},\omega)$ \cite{Damascelli2003, Sobota2021, Hüfner2013}. The latter contains the full spectrum of final states $|  \Psi_f^{N-1} \rangle$ with energy $E_f^{N-1}$, i.\,e. system eigenstates that are accessible after removing one electron with momentum ${\bf k}$ (with operator $c_{\bf k}$) from the $N$-particle ground state 
$| \Psi_0^{N} \rangle$ at energy $E^{N}_0$:
\begin{equation}
A(\mathbf{k},\omega) =
\sum_f
\left|
\langle \Psi_f^{N-1} | c_{\mathbf{k}} | \Psi_0^{N} \rangle
\right|^2
\delta\!\left(\omega - (E_f^{N-1}-E_0^N)\right).
\end{equation}.

In the non-interacting limit, the Green’s function reduces to a sum of simple poles at single-particle band energies $\varepsilon_{n}(\mathbf{k})$,
\begin{equation}
A(\mathbf{k},\omega) \rightarrow \sum_n \delta[\omega - \varepsilon_{n}(\mathbf{k}))],
\end{equation}
and ARPES directly measures the band structure. This is the regime in which the widely used identification “ARPES measures the band structure” is strictly valid.
DFT and its mean-field extensions such as DFT+$U$ implicitly rely on this picture. Even though electronic interactions are partially incorporated through an effective static potential, the excitation spectrum still consists of renormalized single-particle bands. Consequently, within DFT+$U$, ARPES features are interpreted as removing an electron from a Kohn--Sham orbital of the undoped system, and the excitation energy is assumed to correspond to the position of this orbital in the calculated density of states, cf. Fig.~\ref{fig:cartoon}(a).

The above correspondence breaks down in the presence of strong local electron-electron interactions. When these interactions become comparable to or larger than the crystal-field splitting and hybridization energies, a Mott insulating state is formed and in the ground state electrons are completely localized on individual
ions and their nearest-neighbor ligands.
Then, removing an electron in a photoemission no longer creates a single hole in a well-defined Bloch band. Instead, ARPES on Mott insulating NiPS$_3$ can be understood as comprising of two distinct processes.

In the first step, an ARPES photon ejects an electron from the NiS$_6$ cluster. This cluster consists of a Ni$^{2+}$ ion with a $3d^8$ valence shell, hybridized with the $2p^6$ valence shells of six S$^{2-}$ ligands. The removal of the electron in the photoemission results in the formation of several `multiplet' eigenstates corresponding to a NiS$_6$ cluster with Ni$^{3+}$ ($3d^7$), still hybridized with the six S$^{2-}$ ($2p^6$) ligands. This process is pictorially illustrated in the cartoon shown in Fig.~\ref{fig:cartoon}(b).  

In the second step, these multiplet eigenstates of the NiS$_6$ cluster, created by the photoemission process, can propagate through the lattice via exchange mechanisms. This behavior is analogous to previously studied cases, such as the propagation of Ir multiplet eigenstates in Sr$_2$IrO$_4$~\cite{Paerschke2017}, or Ru eigenstates in Ca$_2$RuO$_4$~\cite{Klosinski2020, Revenda2025}. 

While it would be ideal to include both steps in the calculations, it must be emphasized that this is a tremendously challenging task. We note that even understanding the microscopic origin of the spin exchange processes in NiPS$_3$ is rather complex (cf.~\cite{Autieri2022}). For this reason, we are focussing on a detailed calculation of the first step, leaving the second for future work.




Naturally, such a gross simplification of the cluster approach may be concerning. However, it should be stressed that also the complementary DFT$+U$ approach contains severe approximations, as discussed next (see also Fig.~\ref{fig:cartoon}(a)).  


The main limitation of the DFT+$U$ approach in modeling ARPES spectra of Mott insulators is the following:

 (i) In Mott insulators, electrons are localized on transition-metal--ligand clusters as described above. Consequently, the relevant degrees of freedom are interacting electrons on these clusters rather than extended Bloch states.

(ii) {\it Even} for solely noninteracting electrons, ARPES is also sensitive to the final eigenstates of the photoemission, i.\,e. to the doped system (see Fig.~\ref{fig:cartoon}(b)) \cite{Hüfner2013}. 
However, the DFT$+U$ interpretation typically does not differentiate between the following two final states having either three holes in $e_{\rm g}^1 t_{\rm 2g}^2$ (high-spin state, $E_3$ state below)
{\it or}  having three holes in
$e_{\rm g}^0 t_{\rm 2g}^3$ (high-spin state, $E_5$ state below),
even though the difference in energies between these two states is mostly due to crystal field and not due to interactions.
Hence, the DFT$+U$ interpretation just sees {\it one}
crystal field splitting as well as {\it one} charge transfer energy splitting (and possibly also Hund's exchange $J_{H}$) -- i.e. far fewer than the cluster eigenstates shown in Fig.~\ref{Fig. 8} and discussed below.

\subsection{Computational details}
\label{sec:4b}
To model the key ingredients of the correlated electronic structure of NiPS$_3$, we exactly solve the
NiS$_6$ cluster that is modeled
by the following Hamiltonian: 
\begin{equation}
H = H_d + H_L + H_{\text{hyb}} + H_{\text{int}} + H_{\text{SOC}},
\end{equation}
where the individual terms are given by:
\begin{equation}
H_d = \sum_{m,\sigma} \epsilon_d^m\, n_{d,m\sigma},
\end{equation}
\begin{equation}
H_L = \sum_{l,\sigma} \epsilon_L^l\, n_{L,l\sigma},
\end{equation}
\begin{equation}
H_{\text{hyb}} = \sum_{m,l,\sigma} \left( V_{ml}\, d_{m\sigma}^\dagger L_{l\sigma} + \text{h.c.} \right),
\end{equation}


\begin{align}
H_{\text{int}} \!&=\!
\sum_{m<m',\sigma,\sigma'} \! 
U^{(d)}_{mm'}\, n_{d,m\sigma} n_{d,m'\sigma'}\nonumber\\
&- J_H^{(d)} \sum_{m\neq m'} \mathbf{S}_{d,m}\cdot\mathbf{S}_{d,m'}
+\sum_{l<l',\sigma,\sigma'} \! U^{(p)}_{ll'}\, n_{L,l\sigma} n_{L,l'\sigma'}\nonumber\\
&- J_H^{(p)} \sum_{l\neq l'} \mathbf{S}_{L,l}\cdot\mathbf{S}_{L,l'}
+ U_{dp}\sum_{m,l,\sigma,\sigma'} n_{d,m\sigma}\, n_{L,l\sigma'} ,
\end{align}

\begin{equation}
H_{\text{SOC}} = \zeta_i\, \mathbf{L}_d\cdot\mathbf{S}_d.
\end{equation}

Here, $\epsilon_d^m$ and $\epsilon_L^l$ are the on-site energies of Ni~3$d$ (denoted as $d$) and S~3$p$ orbitals (denoted as $L$) with symmetry $m,l \in \{e_{\rm g}, t_{\rm 2g}\}$, while $V_{ml}$ represents the hybridization integral between nickel $d$ and ligand $L$ orbitals. The term $H_{\text{int}}$ describes intra-atomic Coulomb and Hund interactions within the Ni~3$d$ shell, parameterized via Slater integrals ($F^0$, $F^2$, $F^4$). Spin--orbit coupling (SOC) for Ni~3$d$ electrons is included through $\zeta_i$. Note that the trigonal crystal-field distortion in NiPS$_3$ is below 1~meV and was therefore neglected, assuming cubic ($O_h$) symmetry. 

To solve the above model we project
the NiS$_6$ cluster model onto a single impurity Anderson model (SIAM), using ligand field theory~\cite{Eskes1990, Haverkort2012}.
As discussed e.g. in~\cite{He2024} this approach yields results that are essentially identical to those obtained
on a cluster. Within this framework, the Ni$^{2+}$ ion is treated as an interacting impurity (3$d^7$ configuration) hybridized with a bath of S~3$p$ ligand states. The model
captures the essential local physics of the NiS$_6$ cluster, including charge-transfer processes, multiplet formation, and Ni--S hybridization, while remaining computationally tractable. All calculations were carried out within the \textit{exact diagonalization} (ED) framework. The model was solved in the hole representation, consistent with \cite{He2024}. 


\begin{table}[t!]
\centering
\caption{Model parameters used in the ED (SIAM) calculations (all values in eV). 
The origin and interpretation of these parameters are described in detail in He. et al. \cite{He2024}}
\label{tab:SIAMparams}
\begin{tabular}{lll}
\hline
\textbf{Parameter} & \textbf{Description} & \textbf{Value} \\
\hline
$\epsilon_d(e_{\rm g})$ & Ni~3$d$ ($e_{\rm g}$) orbital energy & 0.00 \\
$\epsilon_d(t_{\rm 2g})$ & Ni~3$d$ ($t_{\rm 2 g}$) orbital energy & 0.42 \\
$\epsilon_L(e_{\rm g})$ & Ligand (S~3$p$) $e_{\rm g}$ orbital energy & 6.5 \\
$\epsilon_L(t_{\rm 2g})$ & Ligand (S~3$p$) $t_{\rm 2 g}$ orbital energy & 8.5 \\
$V_{pd\sigma}$ & Ni--S $\sigma$-type hybridization & 0.95 \\
$F^0_{dd}$ & On-site Coulomb integral (Ni~3$d$) & 7.88 \\
$F^2_{dd}$ &  & 10.68 \\
$F^4_{dd}$ &  & 6.68 \\
$F^0_{LL}$ & On-site Coulomb integral (ligand) & 0.46 \\
$F^2_{LL}$ &  & 2.59 \\
$F^4_{LL}$ &  & 1.62 \\
$U_{dL}$ & Ni--ligand intersite Coulomb interaction & 1.0 \\
$\zeta_i$ & Spin--orbit coupling (Ni~3$d$) & 0.083 \\
\hline
\end{tabular}
\end{table}

\subsection{Results and discussion}
\label{sec:4c}
\begin{figure*}[!thbp]
\centering
\includegraphics[width=1\textwidth]{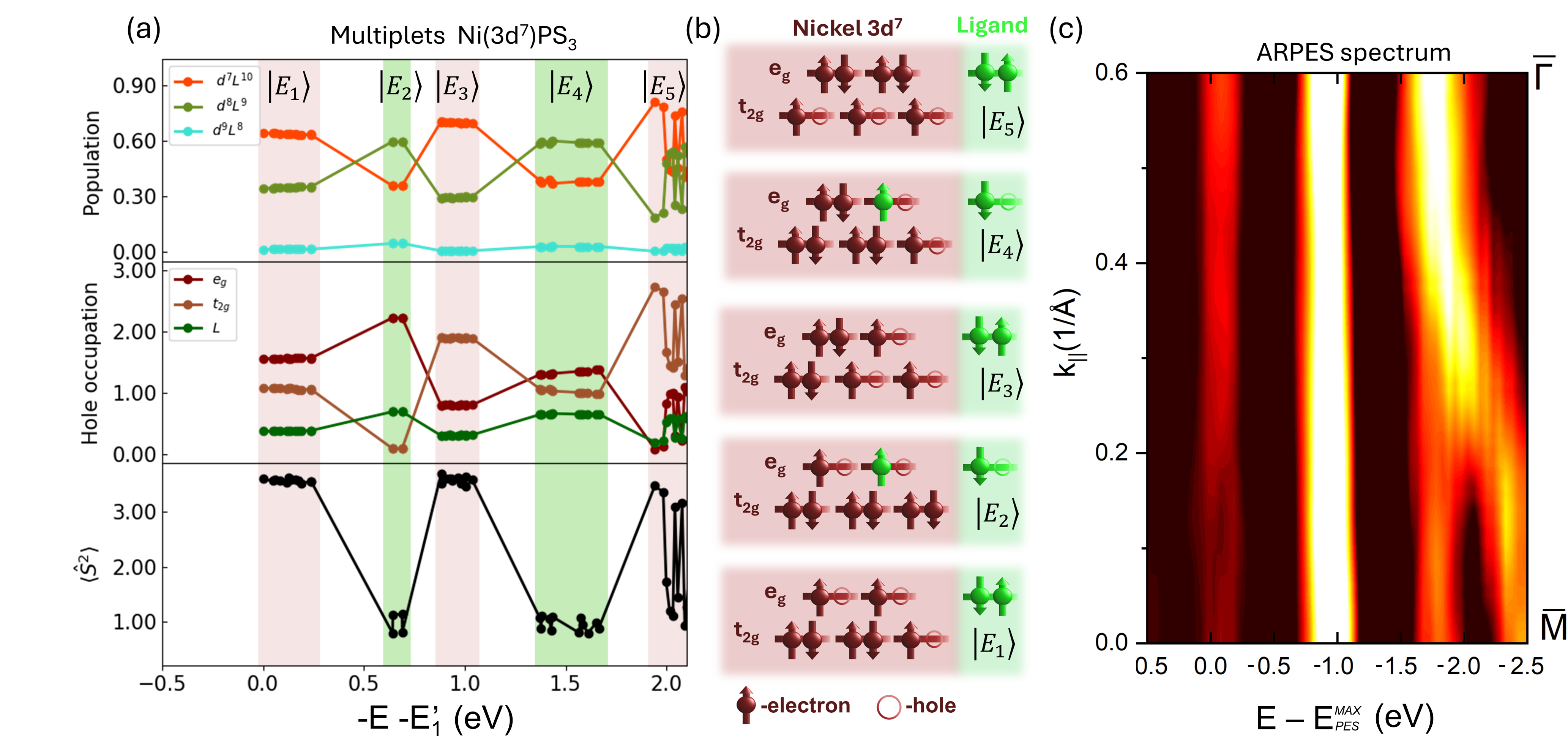}
\vspace{-0.3 cm}
\caption{(a) Results of the ED calculations for Ni(3$d^7$)PS$_3$ (i.e. resolving excited states after electron removal in ARPES), showing the evolution of configuration weights, hole occupations, and total spin moment $\langle S^2 \rangle$ for the lowest multiplet eigenstates $|E_i\rangle$. The ground and low-lying excited states consist mainly of mixed $d^8\underline{L}$ and $d^7$ configurations, with variable ligand-hole contributions reflecting charge-transfer character and Hund’s coupling effects. 
Here, the energy axis is plotted as $-E$ in order to follow the convention. $E'_1$ denotes the lowest-energy state within the $|E_1\rangle$ multiplet manifold.
(b) Schematic representation of all low-energy Ni~3$d^7$ and ligand configurations for selected eigenstates on the cluster, illustrating the redistribution of holes between the Ni $e_{\rm g}$, $t_{\rm 2 g}$, and ligand orbitals across the multiplet manifold.
(c) Experimental ARPES spectrum of NiPS$_3$ along the $\overline{\Gamma}$–$\overline{\mathrm{M}}$ direction, displayed for comparison with the calculated multiplet energies. The similar energy differences between the experimental features and the ED eigenstates implies that the observed additional band structure at $\sim 0$\,eV originates from local Ni–S multiplet excitations. 
Here, $E_{\mathrm{PES}}^{\mathrm{MAX}}$ denotes the energy of the highest-lying band feature in the ARPES spectrum. Note that the ARPES spectrum presented here is from the same experiment as the one presented in Fig.\ref{Fig. 1}(b) -- but the band is shifted to 0 energy, in order to make the comparison with the ED results easier (see text for further details). 
}
\label{Fig. 8}
\end{figure*}

\subsubsection{Results}
The cluster calculations reveal two qualitatively different classes of eigenstates (Fig.~\ref{Fig. 8}a-b). The first class consists of high-spin states nearly without holes on the ligands. These states form three well-separated groups, denoted as $\lvert E_1 \rangle$, $\lvert E_3 \rangle$, and $\lvert E_5 \rangle$, with an energy separation of approximately 1 $\mathrm{eV}$ between successive groups (Fig.~\ref{Fig. 8}a). This energy scale closely matches the separation between the three low-lying features observed in ARPES (Fig.~\ref{Fig. 8}c). Hence, it is tempting to assume that these cluster eigenstates correspond to the experimentally observed three high energy bands. The first group of cluster states predominantly involves holes in the \textit{e}$_g$ orbitals, but also one hole within the \textit{t}$_{2g}$ orbitals, 
whereas the third group has almost exclusively a \textit{t}$_{2g}$ hole character. 

We emphasize that to make the above comparison meaningful, in Fig. \ref{Fig. 8}(c) we shift the energy levels of the experimental ARPES spectrum in such a way that the valence band maximum is located at 0 energy level. This is because in the cluster  ED we merely calculate the energy levels of the NiS$_6$ cluster with three holes -- and we do not take into account the energy cost of adding a hole in the ARPES process, i.e. of going from the $3d^8$ ground state to the $3d^7$ ground state. We note that such an energy cost is anyway poorly defined both in the experimental ARPES spectra and in DFT+U.

The second class of eigenstates comprises low-spin states with the additional hole mostly residing on the ligands. These states form two groups, denoted as $\lvert E_2 \rangle$ and $\lvert E_4 \rangle$, and lie energetically between the final states with the much smaller ligand contribution. Within our previous assignement to the experimental features, these states are missing in the experimental data. One could attribute this to their much stronger inter-cluster hybridisation, which makes them particularly sensitive to the limitations of the single-cluster approximation. We therefore expect that, upon allowing coupling and mixing between neighbouring clusters, the energies and orbital character of these states changes substantially, likely rendering them incoherent or shifting them to a different energy.

An interesting question is how to match the attribution of the upper three experimental bands of Fig.~\ref{Fig. 8}c to cluster states with the previous assignment of two of these bands to the DFT+$U$ results \cite{Pestka2025}. There, the character of the initial single-particle states was predominantly t$_{\rm 2g}$ for the dispersive band at a binding energy of $\sim 2$\,eV and a mixed e$_{\rm g}$ and ligand character for the flat band at $\sim 1$\,eV \cite{Pestka2025}.
Indeed, the $\lvert E_5 \rangle$ state has the largest t$_{\rm 2g}$ character, while $\lvert E_1 \rangle$ and $\lvert E_3 \rangle$ are more e$_{\rm g}$ like. However, the t$_{\rm 2g}$ character of the $\lvert E_5 \rangle$ state  is much stronger than simply removing a single t$_{\rm 2g}$ electron from the ground state t$_{\rm 2g}^6$e$_{\rm g}^2$. A final cluster state with a single removed e$_{\rm g}$ electron does not even exist. This implies that the final states matter most  strongly for the \textit{e}$_g$-derived states. These states are indeed most strongly correlated, since they are largely localized within individual NiS$_6$ clusters, while  \textit{t}$_{2g}$-derived states are comparatively weakly correlated and more mobile.

This qualitative difference, between the \textit{e}$_g$ and the t$_{2g}$ states, has two main microscopic origins. Firstly, the interorbital repulsion within the \textit{e}$_g$ sector is, on average, larger by two Racah parameters~\cite{Ballhausen2009,
Oles2005}
\begin{align}
2B = \frac{2}{49}F^2 - \frac{10}{441}F^4 \approx 0.28~\mathrm{eV}
\end{align}
than the corresponding interaction between the \textit{t}$_{2g}$ and \textit{e}$_g$ sectors, with the absolute value of the latter being approximately 1.1 $\mathrm{eV}$ (in terms of Racah parameters it is $C+2B$ where $C=5 F^4/63 \approx 0.53$ eV~\cite{Ballhausen2009, Oles2005}). Secondly, and far more importantly, when holes occupying \textit{e}$_g$ orbitals attempt to become mobile, they must either overcome the large intraorbital Coulomb repulsion, $U = F^0 \approx 7.88$ $\mathrm{eV}$ (relevant for antiferromagnetic bonds), or are completely blocked by the Pauli principle in the case of ferromagnetic bonds. Together, these effects strongly suppress the mobility of the \textit{e}$_g$-derived states and account for their highly localized character in line with the weak dispersion of these bands observed experimentally.
Hence, we conclude that correlations are much more important for photoemission processes addressing \textit{e}$_g$-type electrons than for any others. This might explain why the single particle \textit{e}$_g$ bands from DFT+$U$ are doubled in the ARPES data pointing to the accessibility of two different groups of final states.

\subsubsection{Discussion}
Altogether, we observe that, depending on the energy scale, the ARPES spectrum can be better described either by the cluster approach \emph{or} by the DFT+$U$ calculations. Specifically, when correlations are strong, \emph{i.e.}, for low-energy states involving the \textit{e}$_g$ electrons, the cluster approach appears to be more appropriate. This is because correlations dominate in this regime (see above), and the electrons are nearly localized on a single cluster.

By contrast, when correlations are relatively weak compared to the kinetic energy, a single-particle band-type description, namely the DFT+$U$ approach, seems adequate. This is the case for the bands with predominantly t$_{\rm 2g}$, sulfur or phosphorus character.

As already mentioned, limiting (or intermediate) case is provided by states with nickel \textit{t}$_{2g}$ character ($|E_3 \rangle$ in the cluster calculations). We find that their onset energy in ARPES is correctly reproduced by the cluster calculations, whereas their dispersion is well captured by DFT+$U$. We suggest that this behavior arises because these quantum many-body states are already quite mobile in reality, and that their interaction can, apparently, be described at the mean-field level. Naturally, more thorough and likely very demanding checks—such as exact diagonalisation of an effective system containing several nickel sites—would be required to verify this hypothesis.

Last but not least, having shown that the cluster approach can be quite useful in partially explaining ARPES results, we now take the opposite perspective and address the following question: what is the general reason that the cluster approach fails here, whereas it works perfectly well for charge-neutral spectroscopies?
Indeed, as shown in Ref.~\cite{He2024}, the cluster model provides an excellent description of ligand-hole--dominated excitations in RIXS. The crucial difference in the present ARPES case lies in the charged nature of the excitation. While optical and RIXS processes create charge-neutral excitations, which in a Mott-insulating background remain largely localized and can propagate only via higher-order exchange processes of order $t^2/U$, ARPES involves the creation of a charged $d^7$ hole. Such charged excitations can, in principle, propagate more efficiently between neighbouring NiS$_6$ clusters via direct hopping processes of order $t$. As a consequence, final states with a substantial ligand-hole character, such as $\lvert E_2 \rangle$ and $\lvert E_4 \rangle$, are expected to be particularly sensitive to inter-cluster coupling. This can lead to strong broadening, redistribution of spectral weight, or shifts in energy that are not captured within the present single-cluster approach, even though the same model successfully describes their charge-neutral counterparts.

\section{Conclusions}
\label{sec:5}
In this work, we have combined detailed $\mu$-ARPES data with complementary theoretical approaches to elucidate the electronic structure of the correlated van der Waals antiferromagnet NiPS$_3$. By systematically confronting experiment with both DFT+$U$ and a multiplet-resolved Single Impurity Anderson Model, we have established a microscopic interpretation of the observed valence-band features that points to the presence of spectral weight in ARPES of NiPS$_3$ beyond a mean-field description.

Within the DFT+$U$ framework, we have shown that all but one of the ARPES-resolved ``bands'' from NiPS$_3$ flakes can be understood in considerable detail. By explicitly analyzing the dependence on the Hubbard $U$ and Hund’s exchange $J_H$, we identified the key mechanisms shaping the single-particle spectrum on the mean-field level: the bonding–antibonding splitting of the Ni $e_{\rm g}$ states driven by strong Ni–S covalency, the spin-dependent splitting of the $t_{\rm 2 g}$ manifold induced by Hund’s coupling, and the crucial role of ligand spin polarization around the magnetic Ni ions. Together, these effects provide a comprehensive description of the ground-state electronic structure at the mean-field level and demonstrate that no additional bands are expected to emerge within DFT+$U$. This analysis allowed us to rule out a mean-field origin of the very weakly dispersive feature observed in ARPES.

To address the discrepancy between ARPES and DFT$+U$, we turned to a cluster-based description, a framework that has previously proven successful in describing many-body excitations in optical and RIXS experiments on NiPS$_3$~\cite{Kang2020, He2024,Song2024, PhysRevLett.131.256504}. Applying this model to the ARPES problem, we demonstrated that electron removal from the NiS$_6$ cluster naturally generates a manifold of multiplet final states with mixed $d^7$ and $d^8\underline{L}$ character. Several of these states appear 
in an energy range of $1-2$\,eV and, hence, provide a natural explanation for the additional ARPES feature absent in DFT+$U$ and located roughly 1\,eV above the valence band maximum. Interestingly, an additional feature appears to be restricted to the e$_{\rm g}$ sector of single particle bands that are naturally most strongly correlated and as such do not appear as a simple single-particle excitation within the multiplet structure claculated by ED. While a one-to-one quantitative correspondence between ED eigenenergies and the experimental ARPES dispersion is not expected due to the single-cluster nature of the model, the semi-quantitative agreement and the richness of the multiplet spectrum strongly indicate a local many-body origin of the observed, additional band-like spectral feature in ARPES. 

Taken together, our results show that a full understanding of NiPS$_3$ requires a combination of complementary theoretical perspectives. DFT+$U$ provides an accurate and indispensable description of the itinerant, band-like aspects of the electronic structure, while the ED of cluster captures correlation-driven, multiplet excitations that are intrinsically beyond any mean-field approach. Only by confronting these two extreme viewpoints---extended single-particle bands on the one hand and fully local many-body cluster states on the other---does the underlying physics become transparent.
Obtaining a fully consistent picture between these two extremes remains one of the central challenges in correlated quantum matter. The present work establishes the ground for future, computationally demanding approaches, such as rather complex extensions of the SCBA approach to fully cover the multiplet physics present here~\cite{Paerschke2017, Klosinski2020, Revenda2025} or the combined DFT and cluster-DMFT calculations.

\section{Acknowledgements and data availability}

We acknowledge financial support from the German Research Foundation (DFG) via the project Mo 858/19-1 and from the German Ministry of Education and Research (Project 05K2022 -ioARPES). 
B.B. acknowledges the support of Humboldt Research Fellowship funded by the Alexander von Humboldt Foundation.
We thank Martin Knupfer for agreeing to use in Fig.~\ref{Fig. 1} the ARPES
data of FePS$_3$ that was previously published as Fig.~1c in~\cite{Koitzsch2023}.
We gratefully acknowledge Polish high-performance computing infrastructure PLGrid (HPC Center: ACK Cyfronet AGH) for providing computer facilities and support within computational grant no. PLG/2025/018673.
Access to computing facilities of the Wroclaw Centre for Networking (WCSS) are gratefully acknowledged.
M.B. acknowledges financial support from the National Science Centre (NCN), Poland under
the grant no. 2024/53/B/ST3/04258. K.W. thanks National Science Center, Poland for financial support 
(grant number 2024/55/B/ST3/03144). A.K.B. and E.L. were supported by the European Commission via the Marie-Sklodowska Curie action Phonsi (H2020-MSCA-ITN-642656). M.R. thanks National Science Center, Poland for financial support, SONATA 19 Grant 2023/51/D/ST11/02588.

ARPES spectra for MnPS$_3$ adapted with permission from J.~Strasdas \textit{et al.}, Nano Letters 23, 10344 (2023)~\cite{Strasdas2023}, copyright © 2023 American Chemical Society. ARPES spectra for
FePS$_3$ adapted from A.~Koitzsch \textit{et al.}, npj Quantum Materials 8, 27 (2023)~\cite{Koitzsch2023}, licensed under CC BY 4.0,
and ARPES spectra for CoPS$_3$ adapted with permission from E.~Voloshina, Y.~Jin, and Y.~Dedkov, Chemical Physics Letters 823, 140511 (2023) \cite{VOLOSHINA2023140511}, copyright © 2023 Elsevier.

\bibliography{references}

\appendix

\section{DFT+U: Bonding–antibonding interpretation of the two \textit{e}$_g$ features in NiPS$_3$}
\label{antibonding}

\begin{figure*}[!thbp]
\centering
\includegraphics[width=1\textwidth]{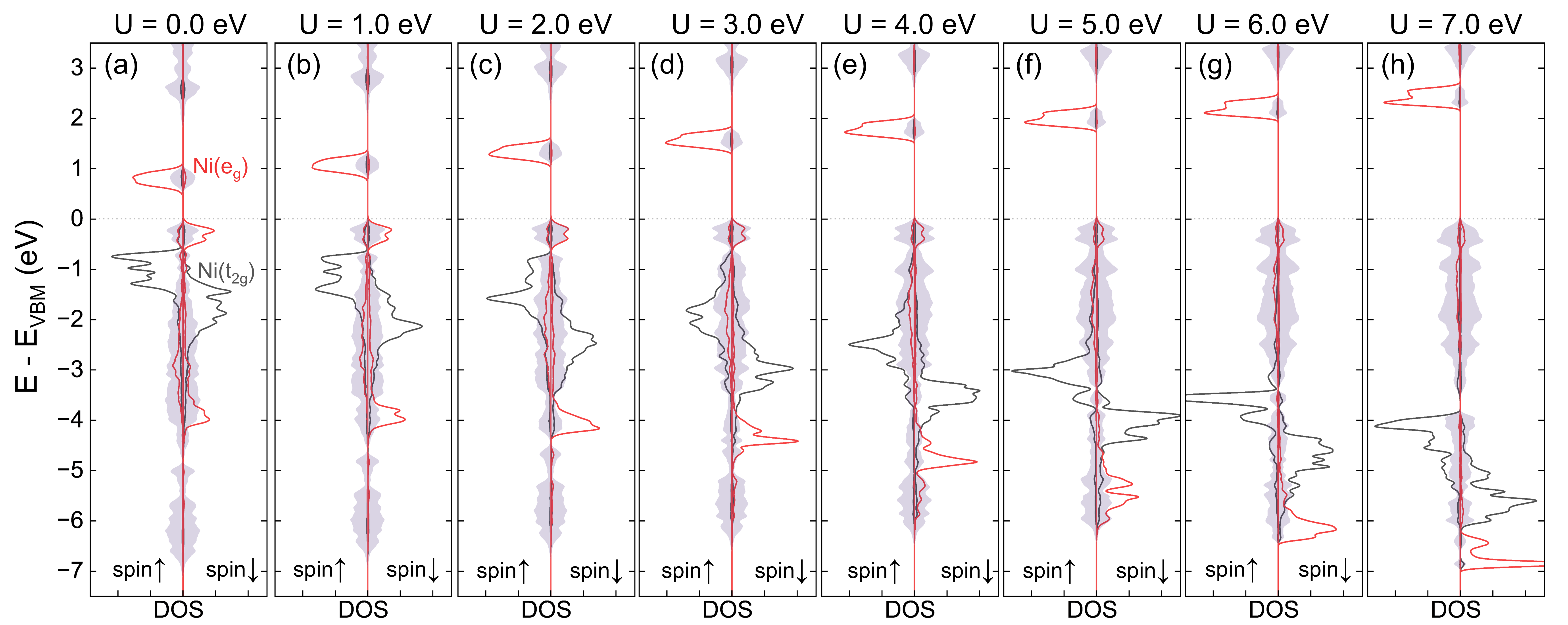}
\vspace{-0.3 cm}
\caption{Evolution of the spin-dependent DOS with an increasing value of Hubbard $U$ in the DFT+U approach. Calculations performed for NiPS$_3$ at $J_H=0.0$ eV, see text for further details. }
\label{fig:s1}
\end{figure*}

\begin{figure}[!thbp]
\centering
\includegraphics[width=0.5\textwidth]{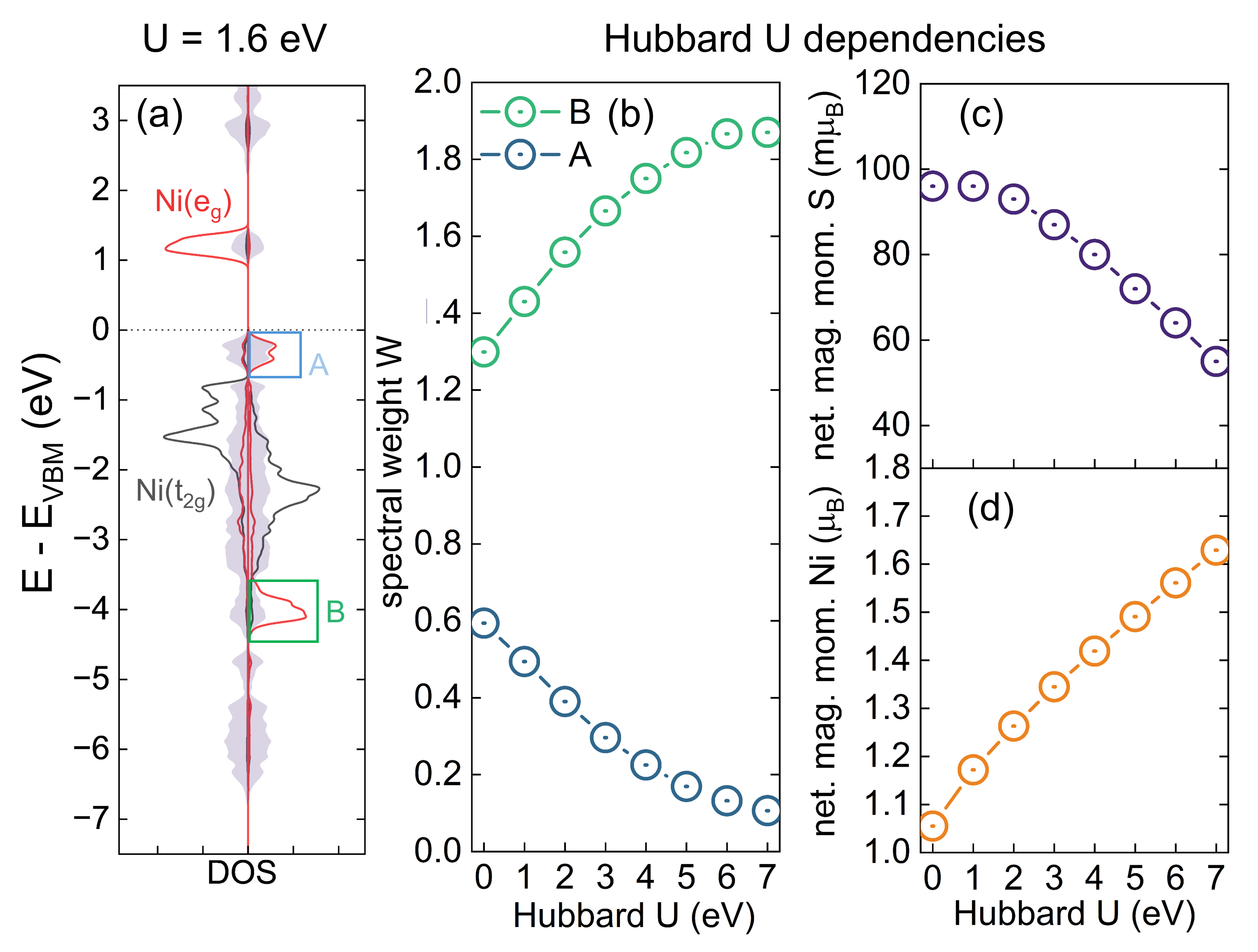}
\vspace{-0.3 cm}
\caption{Hubbard-$U$–driven redistribution of Ni \textit{e}$_g$ spectral weight and its impact on ligand and Ni magnetic moments. 
(a) Spin-resolved Ni $3d$ pDOS at $U=1.6$~eV, highlighting the two \textit{e}$_g$-derived features discussed in the main text: the near-$E_F$ antibonding-like peak (region A) and the deeper-lying bonding-like manifold (region B).  
(b) Integrated spectral weight of regions A and B as a function of the Hubbard interaction $U$. The monotonic transfer of spectral weight from A to B indicates that increasing $U$ progressively suppresses the antibonding \textit{e}$_g$ contribution at the Fermi level while enhancing the occupation of the deeper bonding-like states.  
(c) Net magnetic moment induced on sulfur ligands, showing a systematic decrease with increasing $U$, consistent with reduced Ni–S covalency.  
(d) Net magnetic moment on the Ni site, increasing approximately linearly with $U$ as electronic correlations localize the Ni $3d$ electrons and diminish the ligand contribution to the total magnetic moment.}
\label{fig:s2}
\end{figure}

The two-peak structure observed in the Ni \textit{e}$_g$-projected DOS in Figs.~\ref{fig:s1}–\ref{fig:s2} follows naturally from the hybridization between Ni \textit{e}$_g$ orbitals and ligand $p_\sigma$ orbitals in the NiS$_6$ octahedron. Because the \textit{e}$_g$ orbitals point directly toward the sulfur ligands, they couple through a strong $\sigma$-type hybridization amplitude $V_{dp}$, whereas the \textit{t}$_{2g}$ orbitals, which point between ligands, hybridize only through much weaker $\pi$-type channels. Consequently, only the \textit{e}$_g$ sector satisfies the geometric and energetic conditions required to produce a well-resolved bonding–antibonding pair, while the \textit{t}$_{2g}$ manifold remains comparatively narrow and more weakly hybridized.

To interpret the two-peak structure, we compare the calculated DOS with the simplest two-level model describing hybridization between the nickel $3d (e_{\rm g})$ level with the ligand $2p$ level,
\begin{align} \label{eq:heffmatrix}
H_{\mathrm{eff}} =
\begin{pmatrix}
\epsilon_d & V_{dp} \\
V_{dp} & \epsilon_p
\end{pmatrix}.
\end{align}
Diagonalizing $H_{\mathrm{eff}}$ yields the bonding and antibonding energies
\begin{align}
E_{\pm}
=
\frac{\epsilon_d + \epsilon_p}{2}
\pm
\frac{1}{2}\sqrt{(\epsilon_d - \epsilon_p)^2 + 4 V_{dp}^2}.
\end{align}
For the physically relevant case $\epsilon_d > \epsilon_p$, the corresponding eigenvectors take the form
\begin{align}
\ket{\psi_\pm}
  = \alpha_\pm \ket{d} + \beta_\pm \ket{p_\sigma},
\end{align}
with the analytical weights
\begin{align}
|\beta_\pm|^2 =
\frac12 \left(1 \mp \frac{\epsilon_d - \epsilon_p}{
\sqrt{(\epsilon_d - \epsilon_p)^2 + 4V_{dp}^2}}\right),
\end{align}
and $ | \alpha_\pm|^2= 1- |\beta_\pm|^2 $.
These expressions show that the antibonding state $E_+$ naturally carries more ligand character ($|\beta_+|^2 > 1/2$), whereas the bonding state $E_-$ carries more Ni character ($|\alpha_-|^2 > 1/2$). 


The DFT+$U$ calculations mirror this two-level hybridization physics. The lower \textit{e}$_g$ peak [region B in Fig.~\ref{fig:s2}(a)] corresponds to the bonding solution $E_-$, while the near-$E_F$ peak [region A in Fig.~\ref{fig:s2}(a)] corresponds to the antibonding solution $E_+$. Increasing the Hubbard interaction $U$ effectively shifts the Ni $d$ level upward relative to the ligand level, increasing $\Delta_{dp}(U) = \epsilon_d(U) - \epsilon_p$ and thereby reducing the effective hybridization. In the language of the model above, this suppresses the antibonding weight $|\beta_+|^2$ and strengthens the Ni-centered bonding state. Accordingly, Fig.~\ref{fig:s2}(b) shows the monotonic transfer of integrated spectral weights $W_{A, B}$ from A to B:
\begin{align}
\frac{dW_A}{dU} < 0, \quad
\frac{dW_B}{dU} > 0, \quad
W_A(U) + W_B(U) \approx \text{const}.
\end{align}

The magnetic moments reinforce the same interpretation. Because the antibonding orbital carries substantial ligand character, it polarizes the sulfur atoms. As this state is suppressed with increasing $U$, the ligand moment decreases, exactly as seen in Fig.~\ref{fig:s2}(c). Conversely, the bonding state is predominantly Ni-centered and grows with $U$, leading to the observed increase in the local Ni moment in Fig.~\ref{fig:s2}(d). The opposite $U$-trends of $m_{\mathrm{S}}$ and $m_{\mathrm{Ni}}$ are a direct fingerprint of diminishing $d$–$p$ covalency and simultaneous strengthening of localized Ni character.

Altogether, the symmetry of the Ni–S environment, the electronic configuration of Ni$^{2+}$, the analytic form of the hybridized eigenstates, and the calculated DOS and magnetic moments all point to a consistent picture: the deeper \textit{e}$_g$ peak is the Ni–S bonding state, and the near-$E_F$ \textit{e}$_g$ peak is the corresponding antibonding state. Their evolution with $U$ is fully captured by the simple hybridization model and provides a clear physical explanation for the behavior observed in Figs.~\ref{fig:s1}–\ref{fig:s2}.

The evolution of the antibonding peak across the series MnPS$_3$–FePS$_3$–CoPS$_3$–NiPS$_3$ reflects a systematic change in the relative alignment of the metal $3d$ levels with respect to the ligand $p$ states. In MnPS$_3$ (3$d^5$) and NiPS$_3$ (3$d^8$), the crystal–field splitting is so large that the \textit{e}$_g$ and \textit{t}$_{2g}$ manifolds are energetically well separated. The ligand $p_\sigma$ states lie close in energy to the metal \textit{e}$_g$ orbitals, while the \textit{t}$_{2g}$ orbitals are located significantly deeper and hybridize only weakly. As a result, the two-level hybridization model \eqref{eq:heffmatrix} applies primarily to the \textit{e}$_g$ states, giving rise to a bonding and antibonding pair of predominantly \textit{e}$_g$ symmetry.

In FePS$_3$ (3$d^6$) and CoPS$_3$ (3$d^7$), the situation changes qualitatively. Across the middle of the 3$d$ series, the metal $3d$ levels shift downward in energy and the crystal–field splitting $\Delta_{\mathrm{CF}}=\epsilon_{e_{\rm g}}-\epsilon_{t_{\rm 2g}}$ becomes substantially reduced. As a consequence, both \textit{e}$_g$ and \textit{t}$_{2g}$ orbitals approach energetic resonance with the ligand states, and both types of orbitals participate simultaneously in metal–ligand hybridization. Because a {\it significant} covalency regime
\begin{align}
|\epsilon_d - \epsilon_p| \lesssim 2V
\end{align}
is satisfied for \emph{both} subsets of $d$ orbitals, the antibonding state is no longer of pure \textit{e}$_g$ or pure \textit{t}$_{2g}$ character but instead becomes a mixed combination $c_1 e_{\rm g} + c_2 t_{\rm 2g}$. This mixing is visible directly in the projected DOS, where the antibonding feature contains contributions from both orbital symmetries, and likewise the lower-energy bonding feature acquires a corresponding admixture.

\begin{figure}[!thbp]
\centering
\includegraphics[width=0.45\textwidth]{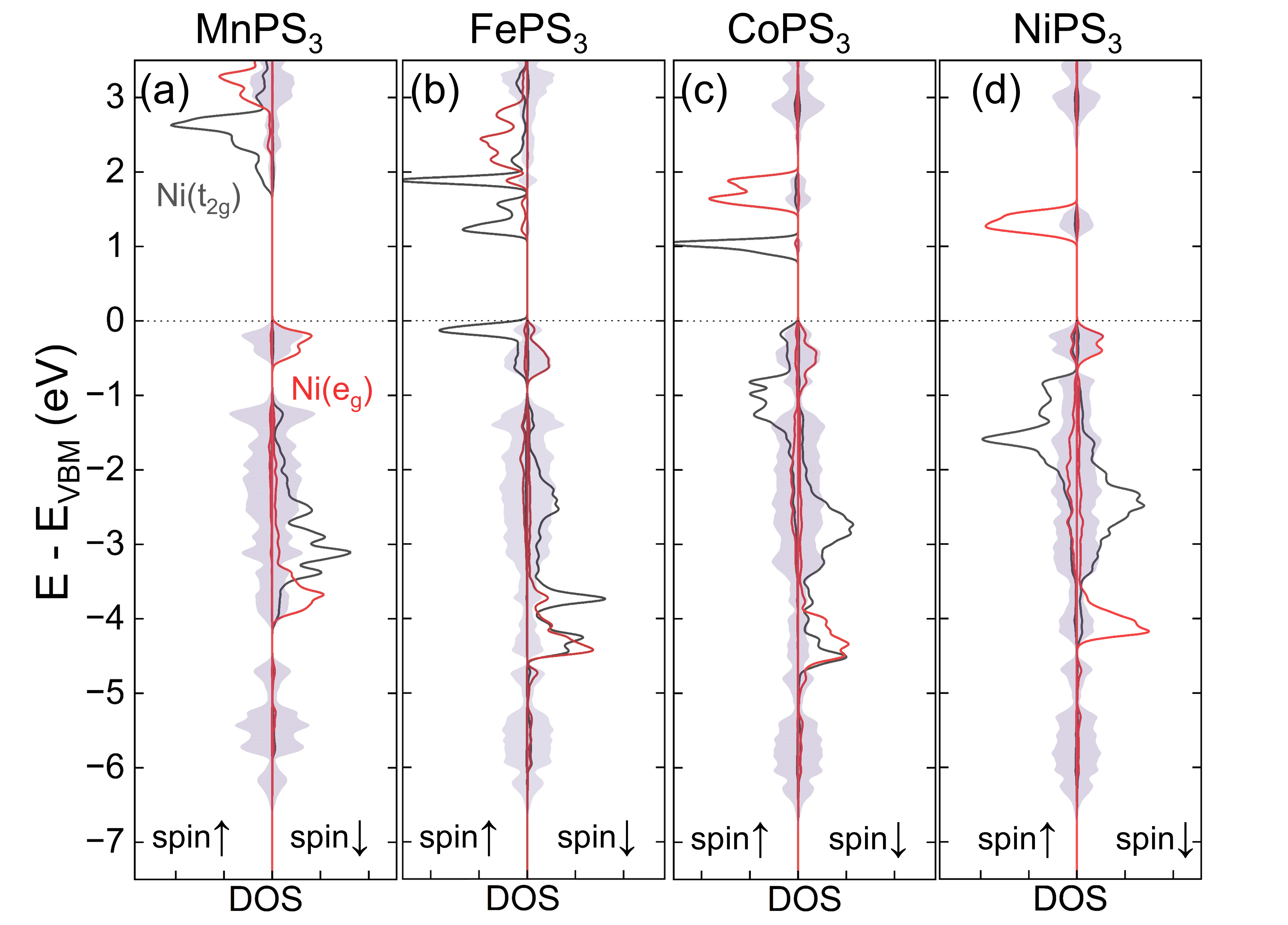}
\vspace{-0.3 cm}
\caption{ Comparison of the transition–metal $3d$ projected DOS across the MPS$_3$ series (M = Mn, Fe, Co, Ni).  
(a–d) Spin-resolved $3d$ projected densities of states for MnPS$_3$, FePS$_3$, CoPS$_3$ and NiPS$_3$, projected onto the \textit{t}$_{2g}$ (black) and \textit{e}$_g$ (red) orbital manifolds. The shaded background shows the total ligand DOS . For MnPS$_3$ and NiPS$_3$ the antibonding feature near the top of the valence band is predominantly of \textit{e}$_g$ character, reflecting the strong $\sigma$-type hybridization between \textit{e}$_g$ and ligand $p_\sigma$ levels. In contrast, for FePS$_3$ and CoPS$_3$ the reduced crystal-field splitting and the different energetic alignment of the metal $d$ levels bring both \textit{t}$_{2g}$ and \textit{e}$_g$ states into resonance with the ligand $p$ manifold, resulting in antibonding features of mixed \textit{e}$_g$–\textit{t}$_{2g}$ character. This systematic evolution across the 3$d$ series supports the hybridization-based interpretation of the bonding–antibonding physics discussed in the main text.}
\label{fig:s3}
\end{figure}

Thus, the origin of the antibonding peak shifts from predominantly \textit{e}$_g$ in MnPS$_3$ and NiPS$_3$ to a mixed \textit{e}$_g$–\textit{t}$_{2g}$ state in FePS$_3$ and CoPS$_3$ because the metal–ligand energy alignment changes across the 3$d$ series. When the ligand level is resonant only with \textit{e}$_g$, the antibonding peak is \textit{e}$_g$-like; when both subsets of $d$ orbitals approach resonance, the antibonding state inevitably acquires a mixed character. This explains the evolution observed in Fig.~\ref{fig:s3}.

\begin{figure*}[!thbp]
\centering
\includegraphics[width=1.0\textwidth]{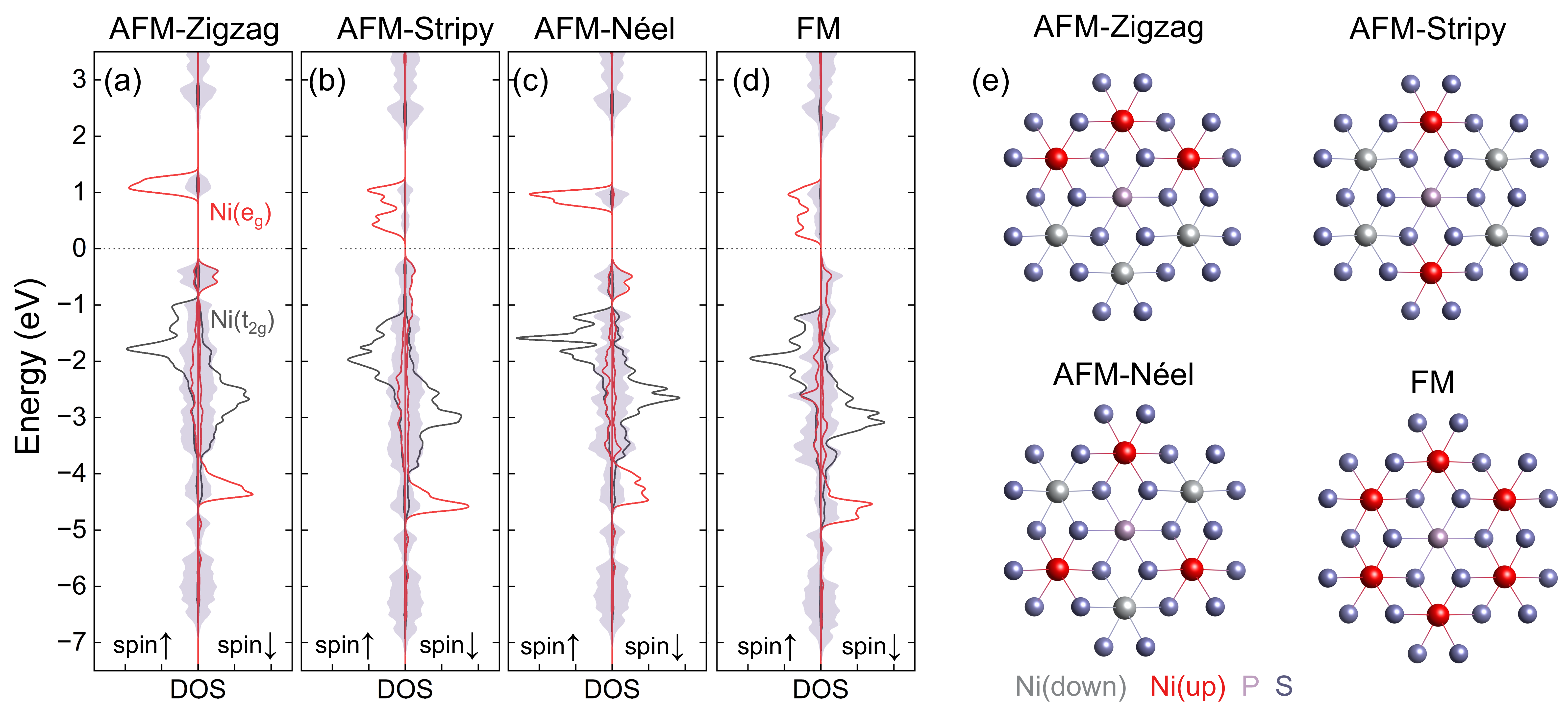}
\vspace{-0.3 cm}
\caption{Robustness of the Ni $3d$ electronic structure of NiPS$_3$ against different magnetic orders. 
(a–d) Spin-resolved Ni $3d$ projected DOS calculated for four competing magnetic configurations: AFM-zigzag (ground state), AFM-stripy, AFM-Néel, and ferromagnetic (FM), all at $U=2.0$~eV. Red and black curves denote the \textit{e}$_g$ and \textit{t}$_{2g}$ contributions, respectively, while the shaded background shows the total ligand DOS. The characteristic splitting of the \textit{e}$_g$ manifold into a near-$E_F$ antibonding feature and a deeper-lying bonding-like band remains essentially unchanged across all magnetic textures.  
(e) Real-space spin configurations corresponding to the four magnetic states, highlighting that the local NiS$_6$ octahedral environment is preserved while only the long-range spin pattern is modified.}
\label{fig:s4}
\end{figure*}

To further test the robustness of the local Ni–S bonding picture developed above, we have calculated the Ni $3d$ projected DOS of NiPS$_3$ for several competing magnetic configurations: AFM-zigzag (the ground state), AFM-stripy, AFM-Néel, and a fully ferromagnetic state (Fig.~\ref{fig:s4}). In all cases the same Hubbard parameters are used, so that only the long-range spin texture is changed while the local NiS$_6$ octahedral environment remains identical.

The resulting spectra demonstrate that the overall structure of the Ni $3d$ DOS is rather insensitive to the choice of magnetic order. In particular, the characteristic splitting of the \textit{e}$_g$ manifold into a deeper bonding-like band and a weakly occupied antibonding peak near the Fermi level persists in all four configurations. Some shifts of spectral weight and changes in spin polarization are observed, in particular for the FM and the AFM-stripy order. However, the generally similar behavior confirms that the bonding–antibonding \textit{e}$_g$ splitting is controlled primarily by local crystal-field and Ni–S hybridization rather than by the spin alignment type.

From the perspective of our cluster-based analysis of the ARPES data, this robustness is important for two reasons. First, it supports the use of a local NiS$_6$ cluster model. The key features of the low-energy electronic structure, in particular, of the upper valence band, are governed by local physics and do not qualitatively depend on whether the spins form a zigzag, stripy, Néel or ferromagnetic pattern. Second, it is consistent with the experimental observation that the additional weakly dispersive feature at the top of the valence band shows little change across the AFM–PM transition.
Together, these results reinforce the conclusion that the relevant hybridized Ni 3\textit{e}$_g$/S 3p states and the associated multiplet excitations are largely local and only weakly affected by long-range magnetic order.

\section{Mean-field Hubbard-Kanamori analysis of the $J_H$ dependence of the $e_{\rm g}$ and $t_{\rm 2g}$ states}
\label{Mean-field}

To clarify the contrasting behavior of the $e_{\rm g}$ and $t_{\rm 2g}$ manifolds upon variation of the Hund’s exchange $J_H$, we analyze the problem within the mean-field formulation of the Hubbard--Kanamori interaction. In the DFT+$U$ framework, the parameters $U$ and $J_H$ do not represent entirely new interactions but rather explicit, orbital-resolved corrections to the averaged Coulomb and exchange effects already contained in the underlying PBE functional. As a consequence, only those interaction terms that generate genuine orbital- or spin-dependent potentials beyond this averaged DFT description can have a physical impact on the single-particle spectrum, while all average contributions must be removed by the double-counting correction. The present mean-field Kanamori analysis is therefore essential for disentangling which parts of the effective single-particle potential correspond to genuine Hund-driven effects and which are artifacts of double counting in the DFT+$U$ scheme.

\paragraph{Double-counting correction in the Liechtenstein DFT+$U$ formalism}

Before turning to the explicit mean-field analysis of the Slater--Kanamori interaction, we clarify the role of the double-counting (DC) correction in the Liechtenstein formulation of DFT+$U$ employed in this work. In this approach, the local Hubbard--Kanamori interaction is added explicitly on top of a conventional density-functional calculation in order to account for orbital-resolved Coulomb and Hund correlations. At the same time, standard DFT already contains an average description of electron--electron interactions through the exchange--correlation functional, expressed in terms of the total charge and spin densities. The purpose of the DC correction is therefore to subtract this average interaction contribution, which would otherwise be counted twice.

In the original work of Liechtenstein \emph{et al.}~\cite{PhysRevB.52.R5467}, the double-counting energy is defined as the interaction energy of an atomic shell with averaged occupancies. In the fully localized limit, it is given explicitly by Eq.~(4) of Ref.~\cite{PhysRevB.52.R5467},
\begin{equation}
E_{\mathrm{DC}}
=
\frac{U}{2}\,N(N-1)
-
\frac{J_H}{2}
\sum_{\sigma}
N_\sigma\bigl(N_\sigma-1\bigr),
\label{eq:dc_energy_liech}
\end{equation}
where $U$ is the intra-orbital Coulomb repulsion, $J_H$ is the Hund’s exchange, $N_\sigma$ denotes the spin-resolved occupation of the correlated shell, and $N=\sum_\sigma N_\sigma$ is the total occupation. Importantly, Eq.~(\ref{eq:dc_energy_liech}) depends only on the total and spin-resolved occupancies and does not resolve individual orbitals, reflecting the fact that it represents an average interaction already contained in the underlying density-functional description.

The corresponding double-counting potential entering the effective single-particle Hamiltonian is obtained by taking the derivative of Eq.~(\ref{eq:dc_energy_liech}) with respect to the orbital occupations. This yields
\begin{equation}
V^{\mathrm{DC}}_{\sigma}
=
\frac{\partial E_{\mathrm{DC}}}{\partial N_\sigma}
=
U\left(N-\tfrac{1}{2}\right)
-
J_H\left(N_\sigma-\tfrac{1}{2}\right),
\label{eq:dc_potential_liech}
\end{equation}
which is independent of the orbital index. Equation~(\ref{eq:dc_potential_liech}) is the form of the DC potential implemented in the Liechtenstein-type DFT+$U$ scheme used in VASP and adopted throughout this work.

Equations~(\ref{eq:dc_energy_liech}) and~(\ref{eq:dc_potential_liech}) make explicit that the DC correction subtracts all Coulomb and Hund contributions that can be expressed purely as functions of $N$ and $N_\sigma$, i.e., orbital-independent average interaction terms already present in standard DFT. Only those components of the Hubbard--Kanamori interaction that generate genuine orbital- or spin-dependent potentials beyond this averaged level survive after double counting. As will be shown below, this property of the DC correction is essential for understanding the contrasting response of the $e_{\rm g}$ and $t_{\rm 2g}$ manifolds to variations of the Hund’s exchange $J_H$.

\paragraph{Mean-field potential from the Slater--Kanamori Hamiltonian}

We consider the density--density part of the local Slater--Kanamori interaction acting on the Ni $d$ shell. Orbitals are labeled by indices $i,j$, and spin by $\sigma=\uparrow,\downarrow$ (with $\bar\sigma$ denoting the opposite spin). The mean-field (Hartree--Fock) single-particle potential acting on orbital $j$ with spin $\sigma$ reads
\begin{equation}
V^{\mathrm{MF}}_{j\sigma}
=
U \langle n_{j\bar\sigma} \rangle
+
(U-2J_H)\sum_{i\neq j}\langle n_{i\bar\sigma} \rangle
+
(U-3J_H)\sum_{i\neq j}\langle n_{i\sigma} \rangle,
\label{eq:kanamori_mf}
\end{equation}
where $\langle n_{i\sigma} \rangle$ denotes the self-consistent orbital occupation.

To make explicit how different orbital subspaces contribute, we decompose the mean-field potential into four channels corresponding to inter- and intra-manifold contributions between the $e_{\rm g}$ and $t_{\rm 2g}$ sectors:
\begin{equation}
V^{\mathrm{MF}}_{j\sigma}
=
V^{(e_{\rm g}\rightarrow e_{\rm g})}_{j\sigma}
+
V^{(t_{\rm 2g}\rightarrow t_{\rm 2g})}_{j\sigma}
+
V^{(e_{\rm g}\rightarrow t_{\rm 2g})}_{j\sigma}
+
V^{(t_{\rm 2g}\rightarrow e_{\rm g})}_{j\sigma},
\label{eq:channels}
\end{equation}
where the notation $X\rightarrow Y$ indicates the contribution originating from subspace $X$ to the potential acting on a state belonging to subspace $Y$ (more precisely, for e.g. $j\in Y$ the corresponding term is obtained by restricting the sums in Eq.~(\ref{eq:kanamori_mf}) to $i\in X$ (with $i\neq j$ where appropriate), while keeping the intra-orbital term $U\langle n_{j\bar\sigma}\rangle$ in the $Y\rightarrow Y$ channel).

In DFT+$U$, this mean-field contribution is supplemented by a double-counting (DC) correction, which subtracts the average interaction already included at the level of the exchange--correlation functional. As already mentioned, in the fully localized limit (FLL) or Liechtenstein formulation, the DC potential depends only on the total occupations, $N_\sigma=\sum_i \langle n_{i\sigma} \rangle$,
and is independent of the orbital index $j$.
The effective potential entering the Kohn--Sham equations is therefore
\begin{equation}
V^{\mathrm{eff}}_{j\sigma}
=
V^{\mathrm{MF}}_{j\sigma}
-
V^{\mathrm{DC}}_{\sigma}.
\end{equation}
As a result, all contributions to $V^{\mathrm{MF}}_{j\sigma}$ that can be expressed solely in terms of total occupations $N$ or $N_\sigma$ (i.e., orbital-independent average shifts) are removed by the double-counting correction, while genuine orbital- or polarization-dependent terms remain.

\paragraph{Inter-manifold contribution $t_{\rm 2g}\rightarrow e_{\rm g}$}

We first consider an orbital $j\in e_{\rm g}$ and evaluate the contribution to its mean-field potential originating from the $t_{\rm 2g}$ sector, i.e., the channel $t_{\rm 2g}\rightarrow e_{\rm g}$ in Eq.~(\ref{eq:channels}). A key property of the system under consideration is that the $t_{\rm 2g}$ states are fully occupied and approximately spin balanced, such that
\begin{equation}
\sum_{i\in t_{\rm 2g}}\langle n_{i\uparrow}\rangle
\approx
\sum_{i\in t_{\rm 2g}}\langle n_{i\downarrow}\rangle
\approx
\frac{N_{t_{\rm 2g}}}{2},
\qquad
N_{t_{\rm 2g}}\approx \text{const}.
\end{equation}
Restricting the sums in Eq.~(\ref{eq:kanamori_mf}) to $i\in t_{\rm 2g}$ then yields
\begin{align}
V^{(t_{\rm 2g}\rightarrow e_{\rm g})}_{e_{\rm g},\sigma}
&=
(U-2J_H)\frac{N_{t_{\rm 2g}}}{2}
+
(U-3J_H)\frac{N_{t_{\rm 2g}}}{2} \nonumber \\
&=
\left(U-\frac{5}{2}J_H\right)N_{t_{\rm 2g}}.
\label{eq:eg_avg}
\end{align}
Importantly, $V^{(t_{\rm 2g}\rightarrow e_{\rm g})}_{e_{\rm g},\sigma}$ is independent of both the spin $\sigma$ and the specific $e_{\rm g}$ orbital. It therefore represents a purely average shift of the $e_{\rm g}$ manifold proportional to the total occupation of the fully filled $t_{\rm 2g}$ shell. Because the double-counting potential subtracts precisely such orbital-independent average contributions, the $J_H$-dependent term in Eq.~(\ref{eq:eg_avg}) is almost canceled in the effective potential acting on $e_{\rm g}$ states. Consequently, after double counting, the remaining $J_H$ dependence of the $e_{\rm g}$ levels is very weak and dominated by much larger energy scales such as the Hubbard $U$, the crystal-field splitting, and Ni--ligand hybridization.

\paragraph{Inter-manifold contribution $e_{\rm g}\rightarrow t_{\rm 2g}$}

We next consider an orbital $j\in t_{\rm 2g}$ and evaluate the contribution originating from the $e_{\rm g}$ sector, i.e., the channel $e_{\rm g}\rightarrow t_{\rm 2g}$ in Eq.~(\ref{eq:channels}). In contrast to the fully occupied $t_{\rm 2g}$ shell, the $e_{\rm g}$ manifold is half-filled and strongly spin polarized. For later convenience, we rewrite the spin-resolved total $e_{\rm g}$ occupancy in terms of its sum and difference. We define the total $e_{\rm g}$ charge and spin polarization as
\begin{equation}
N_{e_{\rm g}}=\sum_{i\in e_{\rm g}}\big(\langle n_{i\uparrow}\rangle+\langle n_{i\downarrow}\rangle\big),
\qquad
M_{e_{\rm g}}=\sum_{i\in e_{\rm g}}\big(\langle n_{i\uparrow}\rangle-\langle n_{i\downarrow}\rangle\big),
\end{equation}
so that, by adding and subtracting these two quantities, the spin-resolved occupations follow algebraically as
\begin{equation}
\sum_{i\in e_{\rm g}}\langle n_{i\uparrow}\rangle=\frac{N_{e_{\rm g}}+M_{e_{\rm g}}}{2},
\qquad
\sum_{i\in e_{\rm g}}\langle n_{i\downarrow}\rangle=\frac{N_{e_{\rm g}}-M_{e_{\rm g}}}{2}.
\end{equation}
Introducing the shorthand $\sigma=\pm1$ for $\uparrow/\downarrow$ spins, this compactly becomes
\begin{equation}
\sum_{i\in e_{\rm g}}\langle n_{i\sigma}\rangle=\frac{N_{e_{\rm g}}}{2}+\sigma\frac{M_{e_{\rm g}}}{2},
\qquad
\sum_{i\in e_{\rm g}}\langle n_{i\bar\sigma}\rangle=\frac{N_{e_{\rm g}}}{2}-\sigma\frac{M_{e_{\rm g}}}{2}.
\end{equation}
Physically, $N_{e_{\rm g}}$ measures the total filling of the $e_{\rm g}$ manifold, while $M_{e_{\rm g}}$ measures its net spin polarization. Substituting these identities into Eq.~(\ref{eq:kanamori_mf}) with the sums restricted to $i\in e_{\rm g}$ yields the $e_{\rm g}\rightarrow t_{\rm 2g}$ contribution:
\begin{align}
V^{(e_{\rm g}\rightarrow t_{\rm 2g})}_{t_{\rm 2g},\sigma}
&=
(U-2J_H)\left(\frac{N_{e_{\rm g}}}{2}-\sigma\frac{M_{e_{\rm g}}}{2}\right)
\nonumber\\
&\quad
+
(U-3J_H)\left(\frac{N_{e_{\rm g}}}{2}+\sigma\frac{M_{e_{\rm g}}}{2}\right)
\nonumber\\
&=
\underbrace{\left(U-\frac{5}{2}J_H\right)N_{e_{\rm g}}}_{\text{average term}}
\;+\;
\underbrace{\sigma\left(-\frac{1}{2}J_H M_{e_{\rm g}}\right)}_{\text{polarization-dependent term}}.
\label{eq:t2g_split}
\end{align}
As in the $t_{\rm 2g}\rightarrow e_{\rm g}$ case, the first term in Eq.~(\ref{eq:t2g_split}) is an orbital-independent average contribution proportional to the total $e_{\rm g}$ occupation and is therefore removed by the double-counting correction. In contrast, the second term depends explicitly on the $e_{\rm g}$ spin polarization $M_{e_{\rm g}}$ and cannot be expressed solely in terms of $N$ or $N_\sigma$. Consequently, this Hund-driven contribution survives the subtraction of $V^{\mathrm{DC}}_\sigma$ and remains in the effective potential acting on the $t_{\rm 2g}$ manifold.

\paragraph{Intra-manifold contributions: $e_{\rm g}\rightarrow e_{\rm g}$ and $t_{\rm 2g}\rightarrow t_{\rm 2g}$}

We now reconsider the intra-manifold channels in Eq.~(\ref{eq:channels}), namely the contributions in which both the source and target orbitals belong to the same symmetry subspace. At the outset, we emphasize an important structural property of the Slater--Kanamori Hamiltonian: all Hund’s-exchange terms involve pairs of \emph{distinct} orbitals ($i\neq j$). Consequently, there is no Hund-driven self-interaction acting within a single orbital. What we refer to as intra-manifold channels therefore corresponds to inter-orbital interactions between symmetry-equivalent orbitals within the same manifold.

We first consider the $e_{\rm g}\rightarrow e_{\rm g}$ channel. For an orbital $j\in e_{\rm g}$, restricting the sums in Eq.~(\ref{eq:kanamori_mf}) to the remaining $e_{\rm g}$ orbital ($j'\in e_{\rm g}$, $j'\neq j$) yields
\begin{align}
V^{(e_{\rm g}\rightarrow e_{\rm g})}_{e_{\rm g},\sigma}
&=
U\langle n_{j\bar\sigma}\rangle
+
(U-2J_H)\langle n_{j'\bar\sigma}\rangle \nonumber \\
&+(U-3J_H)\langle n_{j'\sigma}\rangle.
\label{eq:eg_to_eg}
\end{align}
Writing the spin-resolved occupations of the $e_{\rm g}$ manifold in terms of the total filling $N_{e_{\rm g}}$ and spin polarization $M_{e_{\rm g}}$ (see above) one obtains
\begin{equation}
V^{(e_{\rm g}\rightarrow e_{\rm g})}_{e_{\rm g},\sigma}
=
\frac{3U-5J_H}{4}\,N_{e_{\rm g}}
-
\sigma\frac{U+J_H}{4}\,M_{e_{\rm g}}.
\label{eq:eg_to_eg_result}
\end{equation}
The polarization-dependent term is antisymmetric in spin and therefore produces only a spin splitting between the two spin channels of the $e_{\rm g}$ manifold. However, the energetic location of the $e_{\rm g}$ states is determined by the spin-averaged band position,
\begin{equation}
\bar V_{e_{\rm g}}
=
\frac{1}{2}\left(
V^{(e_{\rm g}\rightarrow e_{\rm g})}_{e_{\rm g},\uparrow}
+
V^{(e_{\rm g}\rightarrow e_{\rm g})}_{e_{\rm g},\downarrow}
\right)
=
\frac{3U-5J_H}{4}\,N_{e_{\rm g}},
\end{equation}
for which the polarization-dependent contribution cancels identically. Thus, the $e_{\rm g}\rightarrow e_{\rm g}$ channel contributes only an orbital-independent average shift proportional to the total $e_{\rm g}$ occupation. Such a contribution depends solely on the total charge and is removed by the double-counting correction. This behavior is physically natural, as Hund’s exchange is not expected to modify the intra-manifold band position, but rather to control spin polarization.

An analogous analysis applies to the $t_{\rm 2g}\rightarrow t_{\rm 2g}$ channel. For an orbital $j\in t_{\rm 2g}$, the corresponding mean-field contribution reads
\begin{align}
V^{(t_{\rm 2g}\rightarrow t_{\rm 2g})}_{t_{\rm 2g},\sigma}
&=
U\langle n_{j\bar\sigma}\rangle
+
(U-2J_H)\sum_{\substack{j'\in t_{\rm 2g}\\ j'\neq j}}\langle n_{j'\bar\sigma}\rangle
\nonumber\\
&\quad
+
(U-3J_H)\sum_{\substack{j'\in t_{\rm 2g}\\ j'\neq j}}\langle n_{j'\sigma}\rangle.
\label{eq:t2g_to_t2g}
\end{align}
Since the $t_{\rm 2g}$ shell is fully occupied and nearly spin balanced, any polarization-dependent terms again enter antisymmetrically in spin and therefore do not affect the spin-averaged energetic position of the $t_{\rm 2g}$ manifold. The resulting contribution reduces to an orbital-independent average shift proportional to $N_{t_{\rm 2g}}$, which is removed by the double-counting correction.

It is instructive to contrast this behavior with the inter-manifold channel $e_{\rm g}\rightarrow t_{\rm 2g}$ discussed above. In that case, the spin polarization of the $e_{\rm g}$ shell enters the effective potential acting on the \emph{different} $t_{\rm 2g}$ manifold, see Eq.~\eqref{eq:t2g_split}. Here the polarization-dependent term does not describe an internal splitting within a single symmetry manifold, but instead controls the \emph{spin-dependent energy separation} between $t_{\rm 2g}^{\uparrow}$ and $t_{\rm 2g}^{\downarrow}$ states. Because the source and target orbitals are not symmetry equivalent, the physically relevant observable is no longer the spin-averaged band position but the spin splitting itself. As a result, no symmetry operation analogous to spin averaging can eliminate this term, and the corresponding $J_H$-dependent contribution survives the double-counting subtraction.

In summary, polarization-dependent terms generated by Hund’s exchange cancel in symmetric intra-manifold channels when one considers the band position as the relevant observable, but remain active in asymmetric inter-manifold channels, where they manifest as genuine spin splittings between different orbital subspaces.

\paragraph{Summary}

The mean-field Kanamori analysis combined with the Liechtenstein double-counting correction provides a clear classification of interaction channels contributing to the effective single-particle potential. All intra-manifold contributions within the $e_{\rm g}$ and $t_{\rm 2g}$ sectors, as well as the inter-manifold contribution from the fully occupied and nearly spin-balanced $t_{\rm 2g}$ shell to the $e_{\rm g}$ states, reduce to orbital-independent average terms that depend only on the total occupations. These contributions are removed by the double-counting correction, explaining why the $e_{\rm g}$ band positions remain essentially insensitive to variations of the Hund’s exchange $J_H$.

In contrast, the inter-manifold contribution originating from the partially filled and spin-polarized $e_{\rm g}$ shell ($e_{\rm g}\rightarrow t_{\rm 2g}$) generates a non-averaged, spin-dependent term in the effective potential acting on the $t_{\rm 2g}$ states. While the accompanying average contribution is again removed by double counting, the remaining polarization-dependent term survives and gives rise to a genuine Hund-driven effective field controlled by $(U-2J_H)$ and $(U-3J_H)$. 

\end{document}

%% file: preamble.tex
\usepackage{amsthm}
\usepackage{mathtools}
\usepackage{physics}
\usepackage{xcolor}
\usepackage{graphicx}
\usepackage[left=23mm,right=13mm,top=35mm,columnsep=15pt]{geometry} 
\usepackage{adjustbox}
\usepackage{placeins}
\usepackage[T1]{fontenc}
\usepackage{csquotes}